\documentclass [11pt]{article}
\usepackage{amsfonts}
\usepackage{graphicx}
\usepackage{amsmath,amssymb, amsfonts, dsfont}
\usepackage{latexsym}
\usepackage[normalem]{ulem}
\usepackage{mathrsfs}
\usepackage{bbold}
\usepackage{color}
\usepackage{slashed}
\usepackage{cancel}
\usepackage{soul} 
\usepackage{setspace} 
\usepackage{comment}
\usepackage{booktabs}
\usepackage{multirow}
\usepackage{tikz-feynman}
\usepackage{pgfplots}
\tikzfeynmanset{compat=1.0.0}
\usepackage{cite}
\usepackage[small]{caption}
\usepackage{epsfig}
\usepackage{slashed}
\usepackage{epstopdf}
\usepackage{latexsym}
\usepackage{graphicx}
\usepackage{subcaption}
\usepackage{tikz}
\usepackage{circuitikz}
\usetikzlibrary{decorations.pathmorphing}
\usetikzlibrary{decorations.markings}

\usepackage{rotating}
\usepackage{bbm}
\usepackage{bbold}
\usepackage{enumitem}

\usepackage{empheq}

\usetikzlibrary{positioning}
\tikzset{
    mybluenode/.style={
        draw=black, circle, minimum width=2cm, inner sep=0pt
        },
    myblacknode/.style={
        circle, inner sep=1pt, fill=black
        },
    }
    
\newcommand*\de{\partial}
\newcommand*\llangle{\langle\!\langle}
\newcommand*\rrangle{\rangle\!\rangle}
\newcommand*\uDi{\underline{\Di}}
\newcommand*\tuDi{\widetilde{\underline{\Di}}}

\newcommand*\al{\alpha}

\newcommand*\ga{\gamma}
\newcommand*\di{\delta}

\newcommand*\vphi{\varphi}

\newcommand*\la{\lambda}

\newcommand*\Di{\Delta}
\newcommand*\Ga{\Gamma}

\newcommand*\RR{\mathbb{R}}

\newcommand*\ZZ{\mathbb{Z}}

\newcommand*\calA{\mathcal{A}}

\newcommand*\calC{\mathcal{C}}

\newcommand*\calG{\mathcal{G}}

\newcommand*\calK{\mathcal{K}}

\newcommand*\calM{\mathcal{M}}

\newcommand*\calO{\mathcal{O}}

\newcommand*\calR{\mathcal{R}}

\def\Res{\mathop{\mathrm{Res}}\limits}

\makeatletter
\def\section{\@startsection {section}{1}{\z@}{-3.5ex plus -1ex minus
 -.2ex}{2.3ex plus .2ex}{\large\bf}}
\def\subsection{\@startsection{subsection}{2}{\z@}{-3.25ex plus -1ex
minus -.2ex}{1.5ex plus .2ex}{\normalsize\bf}}
\makeatother
\makeatletter

\@addtoreset{equation}{section}

\makeatother

\newcommand{\bea}{\begin{equation} \begin{aligned}} \newcommand{\eea}{\end{aligned} \end{equation}}
\def\be{\begin{equation}} \def\ee{\end{equation}}

\usepackage[colorlinks,linkcolor=black,citecolor=blue,urlcolor=blue,linktocpage,pagebackref]{hyperref}

\allowdisplaybreaks

\begin{document}

\thispagestyle{empty}

\begin{flushright}
    \texttt{CERN-TH-2026-121}
\end{flushright}

\begin{center}

	\vspace*{-.6cm}

	\begin{center}

		\vspace*{1.1cm}

	{\centering \Large\textbf{Closing the loop on $\Phi^4$ in AdS$_3$}}

	\end{center}

	\vspace{0.8cm}
	{\bf Dean Carmi$^{a,b,c}$, Riccardo Ciccone$^{a,c}$, Shimon Sukholuski$^{a,d}$}

	\vspace{1.cm}

 ${}^a\!\!$
	{\em Department of Physics and\\ Haifa Research Center for Theoretical
Physics and Astrophysics,\\ University of Haifa, 31905 Haifa, Israel}

	\vspace{.3cm}
${}^b\!\!$
	{\em Department of Mathematics and Physics,\\ University of Haifa at Oranim,
36006 Kiryat Tivon, Israel}

	\vspace{.3cm}
 ${}^c\!\!$
	{\em Theoretical Physics Department,\\ CERN, 1211 Geneva 23, Switzerland}

    \vspace{.3cm}
 ${}^d\!\!$
	{\em Department of Physics,\\ Technion - Israel Institute of Technology,\\ 32000 Haifa, Israel}

\end{center}

\vspace{1cm}

\centerline{\bf Abstract}
\vspace{2 mm}
\begin{quote}
We compute the one-loop correction to the CFT data of all double-trace operators $[\phi\phi]_{n,\ell}$ for a $\Phi^4$ theory in AdS$_3$, for arbitrary values of $n$, $\ell$, and of the scaling dimension $\Delta_\phi>1$. Working in the spectral representation, the $t$-channel one-loop bubble diagram is reduced to a product of spectral integrals dressed by the conformal $6j$ symbol. Both the spectral integrals and the subsequent sums over residues are performed analytically, yielding \emph{finite} closed-form expressions for the anomalous dimensions in terms of higher hypergeometric functions. We discuss the structure of the results, including their large-spin and high-energy behaviors, and show that the anomalous dimensions are completely monotonic in spin.
\end{quote}

\newpage

\tableofcontents

\section{Introduction}
$\lambda\Phi^4$ theory in flat space is arguably the simplest interacting quantum field theory, and its influence in physics has been correspondingly broad. It is the simplest nontrivial example in perturbative QFT and the textbook prototype for understanding renormalization and the organization of UV divergences; its classical action is moreover the Landau-Ginzburg description underlying much of our understanding of continuous phase transitions and critical phenomena. Its perturbative structure, and the spectrum at its interacting fixed points, are understood to remarkable precision. Precisely because of this simplicity and reach, $\lambda\Phi^4$ theory has long served as the canonical testing ground for general ideas.

It is then interesting to ask how much of this familiar story can be made equally explicit in Anti-de Sitter space. Quantum Field Theory in AdS combines local field theory dynamics with the strong kinematic constraints of the AdS isometry group. From the boundary point of view, these isometries act as the conformal group, so that asymptotic observables are naturally encoded by boundary conformal data: operator dimensions, OPE coefficients, and their perturbative corrections in the bulk coupling. This perspective is central in AdS/CFT \cite{Maldacena:1997re,Gubser:1998bc,Witten:1998qj}, but it is useful even when the bulk theory is not gravitational \cite{Callan:1989em,Heemskerk:2009pn,Fitzpatrick:2012cg, Carmi:2018qzm}: AdS provides a maximally symmetric infrared regulator, and the dynamics of a perturbative bulk QFT can be studied precisely and systematically through the lens of boundary conformal correlators. For $\lambda\Phi^4$, as the simplest interacting bulk scalar theory, this is precisely where one should expect the clearest results.

Let us be more precise about what one wants to compute. The observable of interest is the boundary four-point function $\langle\phi\phi\phi\phi\rangle$ of the operator $\phi$ dual to $\Phi$. At zero coupling, this is entirely captured by mean field theory: the $\phi\times\phi$ OPE contains only an infinite tower of double-trace operators $[\phi\phi]_{n,\ell}$, with dimensions $\Delta_{n,\ell}^{(0)} = 2\Delta_\phi + 2n + \ell$ and OPE coefficients fixed by Wick contractions \cite{Heemskerk:2009pn, Fitzpatrick:2010zm}. Switching on interactions makes the problem nontrivial, as it generates anomalous dimensions $\gamma_{n,\ell}$ and corrections to the OPE coefficients $a_{n,\ell}$, organized in a perturbative expansion in $\lambda$. In the bulk, the natural basis for the perturbative expansion is in terms of Witten diagrams. 
However, the direct computation of Witten diagrams is notoriously challenging already at lowest order. 
This has motivated several complementary approaches to perturbative AdS dynamics: Mellin space methods \cite{Mack:2009mi, Penedones:2010ue, Fitzpatrick:2010zm,Fitzpatrick:2011hu, Rastelli:2016nze,Yuan:2017vgp,Yuan:2018qva,Huang:2023ppy,Ghosh:2019lsx}, analytic bootstrap techniques \cite{Fitzpatrick:2011dm,Costa:2012cb,Fitzpatrick:2012yx, Komargodski:2012ek,Aharony:2016dwx,Alday:2015eya,Kaviraj:2015cxa,Kaviraj:2015xsa,Alday:2016njk, Caron-Huot:2017vep, Alday:2017vkk,Aprile:2017bgs,Albayrak:2019gnz,Huang:2023oxf,Bissi:2025tci}, crossing kernels, $6j$ symbols, and spectral representation \cite{Giombi:2017hpr,Sleight:2017fpc,Sleight:2018epi,Sleight:2018ryu,Carmi:2018qzm, Liu:2018jhs,Meltzer:2019nbs,Ponomarev:2019ltz,Carmi:2019ocp,Carmi:2021dsn,Carmi:2024tzj,Albayrak:2020rxh,Dujava:2025php}, and direct computations in position \cite{Akhmedov:2018lkp,Cacciatori:2024zbe,Bertan:2018afl,Bertan:2018khc,Heckelbacher:2022fbx,Xiao:2026prw,Ciccone:2024guw,Ciccone:2025dqx,Ankur:2026ylr} and momentum space \cite{Raju:2010by,Raju:2011mp,Albayrak:2018tam,Albayrak:2019asr,Albayrak:2020bso, Albayrak:2020isk,Albayrak:2023kfk}.  
The first three approaches make different aspects of the problem manifest: Mellin space emphasizes the analogy with flat-space scattering amplitudes; the analytic bootstrap organizes CFT data through crossing symmetry and the large-spin expansion; crossing kernels and the spectral representation expose the representation-theoretic structure underlying Witten diagrams and organize the integrand in a basis where CFT data can be read off directly. Together, these methods have yielded loop-level results in a number of theories and special cases, yet general closed-form expressions remain elusive, a gap that is especially striking for $\lambda\Phi^4$ theory, where existing results are confined either to special values of the external scaling dimension or to the leading Regge trajectory \cite{Bertan:2018afl,Bertan:2018khc,Heckelbacher:2022fbx,Xiao:2026prw}.

In this work we compute the one-loop correction to the CFT data of all double-trace operators $[\phi\phi]_{n,\ell}$ in $\lambda\Phi^4$ theory in AdS$_3$, for arbitrary $n$, $\ell$, and external dimension $\Delta_\phi > 1$. We focus on AdS$_3$ as the simplest setting in which the computation can be carried out in full generality. Our strategy is direct, but uses conformal symmetry at every stage. We write the one-loop bubble in spectral form and reduce the problem to the extraction of boundary conformal data. The $s$-channel bubble is treated by closing the spectral contour for the bubble function and reading off poles at double-trace locations, as done in \cite{Carmi:2018qzm,Carmi:2019ocp}. For the crossed-channel bubbles, we use the conformal $6j$ symbol to decompose $t$-channel partial waves into $s$-channel partial waves \cite{Liu:2018jhs,Meltzer:2019nbs}. After evaluating the relevant residues, the result first appears as an infinite sum over intermediate states. We show that this sum can be performed analytically, yielding finite closed-form expressions for the anomalous dimensions in terms of very-well-poised hypergeometric functions ${}_{7}F_6(1)$. In the special case of half-integer scaling dimension, it further reduces to a finite sum of terminating $_4F_3(1)$. To the best of our knowledge, such general closed-form expressions have not appeared in the literature before.

Beyond the explicit results, we analyze several structural properties of the anomalous dimensions. The large-spin asymptotics of $\gamma^{(2)}_{n,\ell}$ reproduces the predictions of the lightcone bootstrap, with subleading corrections in even powers of the conformal spin, and we compute the behavior in the Regge and bulk-point limits. Intriguingly, we find that the double-trace anomalous dimensions are completely monotonic functions of $\ell$ for all values of $n$  and $\Delta_\phi > 1$, namely they satisfy $(-1)^r \partial_\ell^r \gamma_{n,\ell}^{(2)}<0$, generalizing the pattern of negativity, monotonicity, and convexity \cite{Komargodski:2012ek} to all orders in derivatives with respect to spin. We also clarify the origin and the validity of the so-called ``derivative relations'' \cite{Heemskerk:2009pn,Fitzpatrick:2011dm} connecting perturbative corrections to OPE coefficients and anomalous dimensions from the viewpoint of spectral representation.

The rest of this paper is organized as follows. In section \ref{sec:lambdaphi4} we set up $\lambda\Phi^4$ theory in AdS$_{d+1}$, introduce the spectral representation, and analyze the tree-level corrections to OPE data for generic $d$. In section \ref{sec:oneloop} we specialize to AdS$_3$ and compute the one-loop corrections: in subsection \ref{subsec:sbubble} the $s$-channel bubble is treated by closing the spectral contour, while in subsection \ref{subsec:tbubble} the $t$- and $u$-channel bubbles are handled via the conformal $6j$ symbol, yielding the closed-form ${}_{7}F_6(1)$ expressions for the anomalous dimensions. Section \ref{sec:specialized} records several specializations and limits: reductions at half-integer $\Delta_\phi$ in \ref{subsec:speecialdelta}, large-spin asymptotics in \ref{subsec:largespin}, and Regge and bulk-point limits in \ref{subsec:highenergy}. In section \ref{sec:discussion} we discuss the structure of the results, including the scheme ambiguity and its resolution  in subsection \ref{sec:schemeindependent}, the derivative relation in subsection \ref{subsec:derivativerel}, and the complete monotonicity properties in subsection \ref{subsec:monotonicity}. Appendix \ref{sec:proof} contains a sketch of the proof of the closed-form representation.

\section{$\lambda\Phi^4$ theory in AdS$_{d+1}$}\label{sec:lambdaphi4}
The action for $\lambda\Phi^4$ theory in Euclidean AdS$_{d+1}$ is given by
\begin{equation}\label{eq:phi4}
    S=\int d^{d+1}x\sqrt{g}\, \left(\frac12(\nabla_\mu\Phi)^2+\frac{1}2M^2\Phi^2+\frac1{4!}\lambda\Phi^4\right)\,,
\end{equation}
with the background metric being, in Poincaré coordinates,
\begin{equation}
    ds^2=\frac{L^2}{z^2}(dz^2+dx_i dx^i)\,,\qquad i=1,\dots,d\,.
\end{equation}
In what follows we will set $L=1$ for simplicity.

At the boundary of AdS the field $\Phi$ can have two possible asymptotic behaviors, $\Phi\sim z^{\Delta_\phi}(\phi+\calO(z^2))+z^{d-\Delta_\phi}(\vphi+\calO(z^2))$. In the absence of interactions, the scaling dimension $\Delta_\phi$ corresponds to the one determined by the parameter $M^2$ as $M^2=\Delta_\phi(\Delta_\phi-d)$.
In order to have a well-defined variational problem that leads to a stable boundary condition, one must freeze either $\varphi$ (standard quantization, or Dirichlet) or $\phi$ (alternative quantization, or Neumann).
In an interacting theory, we cannot repeat this reasoning as the bare value of $M^2$ is not a physical quantity: what remains physical is the scaling dimension of the boundary operator $\Delta_\phi$, which we keep fixed as renormalization condition. Notice also that we should include in the action \eqref{eq:phi4} the appropriate counterterms to cancel bulk UV divergences. In the following we will work in standard quantization, that is $\Di_\phi>d/2$.

Given these premises, the bulk-to-bulk propagator of $\Phi$ is given by
\begin{equation}\label{eq:genericbulktobulk}
    G_{\Delta_\phi}(X,Y)=\frac{\Gamma(\Delta_\phi)}{2 \pi^{\frac{d}{2}} \Gamma\left(\Delta_\phi+1-\frac{d}{2}\right)}(2 u)^{-\Delta_\phi}{ }_2 F_1\left(\Delta_\phi, \Delta_\phi+\frac{1-d}{2}, 2 \Delta_\phi-d+1,-\frac{2}{u}\right) \,,
\end{equation}
where $u=-1-X\cdot Y$ is the chordal distance between the two points.
Here and throughout the paper we adopt embedding space formalism for AdS$_{d+1}$ and for the boundary conformal theory \cite{Costa:2011mg,Costa:2014kfa}, where coordinates are described by points on a hyperboloid (and null rays, respectively) in $\RR^{d+1,1}$. The relation between embedding space coordinates and Poincaré coordinates is
\begin{equation}
    X^A=\frac1{z}\left(\frac{1+(\vec{x}^2+z^2)}2,\frac{1-(\vec{x}^2+z^2)}2,\vec{x}\right)\,,
\end{equation}
where $X\cdot X=-1$ with metric $(-,+,\dots,+)$. Points on the boundary are described by null rays in $\RR^{d+1,1}$, that is $P\sim \lambda P$, $P^2=0$. We choose the frame in which they are parametrized by 
\begin{equation}
    P^A=\left(\frac{1+\vec{x}^2}2,\frac{1-\vec{x}^2}2,\vec{x}\right)\,.
\end{equation}

Taking the boundary limit of one of the two points in \eqref{eq:genericbulktobulk}, we obtain the bulk-to-boundary propagator
\begin{equation}
    K_{\Delta_\phi}(X,P)=\frac{\calC_{\Delta_\phi}}{(-2P\cdot X)^{\Delta_\phi}}\,,
\end{equation}
with the two-point coefficient being
\begin{equation}
    \calC_{\Delta_\phi}=\frac{\Ga(\Di_\phi)}{2\pi^{d/2}\Ga({\Di_\phi-d/2+1})}\,.
\end{equation}

At the boundary, the observable of interest is the four-point function of the operator $\phi$,
\begin{equation}\label{eq:generic4pt}
   \calA(P_i)\equiv  \langle \phi(P_1)\phi(P_2)\phi(P_3)\phi(P_4)\rangle = \frac{\calC_{\Di_\phi}^2}{(P_{12})^{\Delta_\phi}(P_{34})^{\Delta_\phi}}\calG(z,\bar{z})\,,
\end{equation}
We explicitly factor out the normalization of the operators, so that the function $\calG$ is independent of the choice of the latter. Here, $z,\bar z$ are the conformal cross ratios
\begin{equation}
    z\bar z = \frac{P_{12} P_{34}}{P_{13}P_{24}}\,,\qquad (1-z)(1-\bar z)=\frac{P_{14}P_{23}}{P_{13}P_{24}}\,,
\end{equation}
and $P_{ij}\equiv -2P_i\cdot P_j$.

In general, $\calG(z,\bar{z})$ admits an expansion in terms of conformal blocks, 
\begin{equation}\label{eq:genericblockexp}
    \calG(z,\bar{z})=1+\sum_{\calO} a_{\calO} \calK_{\Delta_\calO, \ell_\calO}(z,\bar z)\,,
\end{equation}
where the sum runs over all operators $\calO$ different from the identity appearing in the $\phi\times \phi$ OPE, and $a_\calO= c_{\phi\phi \calO} c^{\phi\phi\calO}\geq0$ are the squared OPE coefficients. Only even spins $\ell_\calO$ contribute when the theory has no internal symmetry.
Since the AdS background is fixed and gravity is not dynamical, the boundary data considered here does not assume a conserved stress tensor in the $\phi\times \phi$ OPE.

We set up a perturbative expansion in $\la$,
\begin{equation}
    \calA(P_i)=\calA_{\rm MFT}(P_i) + \la\, \calA^{(1)}(P_i)+ \la^2\,\calA^{(2)}(P_i)+\dots\,,
\end{equation}
or, equivalently,
\begin{equation}
    \calG(z,\bar{z})=\calG_{\rm MFT}(z,\bar z)+\la\, \calG^{(1) }(z,\bar z)+\la^2\,\calG^{(2)}(z,\bar z)+\dots\,.
\end{equation}
The leading order $\calG_{\rm MFT}$ is just the mean field theory correlator,
\begin{equation}
    \calG_{\rm MFT}(z,\bar z)=1+(z\bar z)^{\Delta_\phi}+ \left(\frac{z\bar z}{(1-z)(1-\bar z)}\right)^{\Delta_\phi}\,.
\end{equation}
We can expand $\calG_{\rm MFT}$ in terms of conformal blocks,
\begin{equation}
    \calG_{\rm MFT}=1+\sum_{n=0}^\infty\sum_{\substack{\ell=0}}^\infty\left(1+(-1)^\ell\right)c_{n,\ell}^2\calK_{2\Di_\phi+2n+\ell,\ell}(z,\bar{z})\,,
\end{equation}
with
\begin{equation}
c_{n, \ell}^2=\frac{(-1)^{\ell}\left[\left(\Delta_\phi-\frac{d}{2}+1\right)_n(\Delta_\phi)_{\ell+n}\right]^2}{\ell!n!\left(\ell+\frac{d}{2}\right)_n(2 \Delta_\phi+n-d+1)_n(2 \Delta_\phi+2 n+\ell-1)_{\ell}\left(2 \Delta_\phi+n+\ell-\frac{d}{2}\right)_n}\,.
\end{equation}
The sum runs over the double-trace operators $[\phi\phi]_{n,\ell}$, which in the free theory saturate the $\phi\times \phi$ OPE.
From the above, it is manifest that in MFT they have scaling dimension $\Delta^{(0)}_{n,\ell}=2\Delta_\phi+2n+\ell$ and squared OPE coefficient $a_{n,\ell}^{(0)}=\left(1+(-1)^\ell\right)c_{n,\ell}^2$. In particular, for $d=2$,
\begin{equation}
    a_{n,\ell}^{(0)}\underset{d=2}{=}\bigl(1+(-1)^\ell\bigr)
\frac{(\Delta_\phi)_n^2}
{n!\,(2\Delta_\phi+n-1)_n}
\frac{(\Delta_\phi)_{n+\ell}^2}
{(n+\ell)!\,(2\Delta_\phi+n+\ell-1)_{n+\ell}}\,.
\end{equation}

\subsection{Tree-level corrections to OPE data}

The leading correction to \eqref{eq:generic4pt} with respect to its MFT value corresponds to a contact interaction in the bulk,
\begin{align}
    \calA^{(1)}(P_i)&=- \int dX\, K_{\Delta_\phi}(P_1,X) K_{\Delta_\phi}(P_2,X) K_{\Delta_\phi}(P_3,X) K_{\Delta_\phi}(P_4,X)\,.
\end{align}
We split the bulk integral by introducing the spectral representation \cite{Penedones:2010ue},
\begin{equation}
     \di(X,Y)=-i\int d\underline{\Di}\,\Omega_{\underline{\Di}}(X,Y)\,,
\end{equation}
where the integral runs over the principal series $\frac d2+i\RR$, and $\Omega_{\uDi}(X,Y)$ is the AdS scalar harmonic of spectral parameter $\uDi$,
\begin{equation}
    \left(-\nabla^2_X+\uDi(\uDi-d)\right)\Omega_{\uDi}(X,Y)=0\,.
\end{equation}
The latter admits a representation as the convolution of bulk-to-boundary propagators,
\begin{equation}\label{eq:splitrep}
    \Omega_{\uDi}(X,Y)=-\frac{(\uDi-\tuDi)^2}{4\pi}\int_\de dP\ K_{\uDi}(X,P)K_{\tuDi}(Y,P)\,,
\end{equation}
where $\tuDi=d-\uDi$.
For instance, the scalar propagator admits the spectral representation
\begin{equation}\label{eq:spectralrepprop}
    G_{\Di_\phi}(X,Y)=i\int {d\underline\Di}\frac{1}{(\underline\Di- d/2)^2-(\Di_\phi- d/2)^2}\Omega_{\uDi}(X,Y)\,,
\end{equation}
and in general any AdS bilocal scalar function can be decomposed in such a basis.

After introducing the spectral representation, one has
\begin{align}
    \calA^{(1)}(P_i)&=i \int d\uDi\,  I_{\uDi}(P_i)\,,
\end{align}
with \cite{Carmi:2018qzm}
\begin{equation}
\begin{aligned}
I_{\uDi}(P_i)&\equiv \int dX\, dY\, \Omega_{\underline{\Di}}(X,Y)K_{\Delta_\phi}(P_1,X) K_{\Delta_\phi}(P_2,X) K_{\Delta_\phi}(P_3,Y) K_{\Delta_\phi}(P_4,Y)\\
&=-\frac{(\uDi-\tuDi)^2}{4\pi} b_{\Di_\phi\Di_\phi\uDi}b_{\tuDi \Di_\phi\Di_\phi}\Psi_{\uDi,0}^{s}(P_i)\,,
\end{aligned}
\end{equation}
where we have used \eqref{eq:splitrep} and
\begin{equation}
    \int dX\, K_{\Delta_1}(X,P_1)K_{\Delta_2}(X,P_2)K_{\Delta_3}(X,P_3)=b_{\Delta_1,\Delta_2,\Delta_3}\llangle \calO_{\Delta_1}(P_1)\calO_{\Delta_2}(P_2)\calO_{\Di_3}(P_3) \rrangle\,,
\end{equation}
where $\llangle \calO_{\Delta_1}(P_1)\calO_{\Delta_2}(P_2)\calO_{\Di_3}(P_3) \rrangle$ is the conformal three-point structure, and the three-point coefficient reads 
\begin{equation}
b_{\Delta_1,\Delta_2,\Delta_3}
=
\tfrac{\pi^{\frac d2} \mathcal C_{\Delta_1}\mathcal C_{\Delta_2}\mathcal C_{\Delta_3}}{2\Gamma(\Delta_1)\Gamma(\Delta_2)\Gamma(\Delta_3)}
\Gamma\big(\tfrac{\Delta_1+\Delta_2+\Delta_3-d}{2}\big)
\Gamma\big(\tfrac{\Delta_1+\Delta_2-\Delta_3}{2}\big)
\Gamma\big(\tfrac{\Delta_2+\Delta_3-\Delta_1}{2}\big)
\Gamma\big(\tfrac{\Delta_1+\Delta_3-\Delta_2}{2}\big)\,.
\end{equation}
The function $\Psi_{\uDi,0}^s(P_i)$ is the $s$-channel scalar conformal partial wave of dimension $\uDi$, which is defined as the convolution of two three-point structures. Allowing for the possibility of a generic spin $J$, it is given by
\begin{equation}
\begin{aligned}
        \Psi_{\uDi,J}^s(P_i)&\equiv \int_{\de} dP_0\ \llangle \phi(P_1)\phi(P_2)\calO_{\uDi,J}(P_0)_{A_1\dots A_J} \rrangle\llangle \widetilde\calO_{\widetilde{\uDi},J}(P_0)^{A_1\dots A_J}\phi(P_3)\phi(P_4)\rrangle\\
        &=\frac1{(P_{12})^{\Di_\phi}(P_{34})^{\Di_\phi}}\left(K_{\tuDi,J}\calK_{\uDi,J}(z,\bar z)+ K_{\uDi,J}\calK_{\tuDi,J}(z,\bar{z})\right)\,,
\end{aligned}
\end{equation}
with
\begin{equation}
    K_{\Delta,J}=\frac{\pi^{\frac{d}{2}} \Gamma\left(\Delta-\frac{d}{2}\right) \Gamma\left(\Delta+J-1\right) \Gamma\left(\frac{\widetilde{\Delta}+J}{2}\right)^2}{\Gamma\left(\Delta-1\right) \Gamma\left(d-\Delta+J\right)  \Gamma\left(\frac{\Delta+J}{2}\right)^2}\,.
\end{equation}
All together, one finds
\begin{equation}
    \begin{aligned}\label{eq:spectralG1}
        \calG^{(1)}(z,\bar z)&=\int \frac{d\uDi}{2\pi i}\frac{(\uDi-\tuDi)^2}{\calC_{\Di_\phi}^2}b_{\Di_\phi\Di_\phi\uDi}b_{\tuDi \Di_\phi\Di_\phi}K_{\tuDi,0}\calK_{\uDi,0}\,,\\
    \end{aligned}
\end{equation}
which has double poles at scalar double-trace locations $\uDi=\Di_{n,0}^{(0)}$ coming from the three-point coefficients. Closing the contour to the right and comparing with 
\begin{equation}
    \calG^{(1)}(z,\bar z)=\sum_{n=0}^\infty\sum_{\substack{\ell=0\\\ell \text{ even}}}^\infty\left(a_{n,\ell}^{(1)}+\frac12a_{n,\ell}^{(0)}\ga_{n,\ell}^{(1)}\frac\de{\de n}\right)\calK_{\Delta_{n,\ell}^{(0)},\ell}(z,\bar{z})\,,
\end{equation}
we can extract $\gamma_{n,\ell}^{(1)}$ and $a_{n,\ell}^{(1)}$ from the residues at each double-trace pole \cite{Heemskerk:2009pn,Aharony:2016dwx}, 
\begin{align}\label{eq:gamma1genericd}
    \gamma_{n,\ell}^{(1)}&= \frac{\delta_{\ell,0}}{2^{d+2}\pi^{d/2}\Gamma\!\left(\frac d2\right)}
\frac{
\Gamma\!\left(\frac d2+n\right)
\Gamma(\Delta_\phi+n)
\Gamma\!\left(\Delta_\phi+n+\frac12-\frac d2\right)
\Gamma\!\left(2\Delta_\phi+n-\frac d2\right)
}{
n!\,
\Gamma\!\left(\Delta_\phi+n+\frac12\right)
\Gamma\!\left(\Delta_\phi+n+1-\frac d2\right)
\Gamma\!\left(2\Delta_\phi+n+1-d\right)
}\,,
\end{align}
\begin{equation}
    \label{eq:treederivativeformula} a_{n,\ell}^{(1)}=\frac12\frac{\de}{\de n}\left(a_{n,\ell}^{(0)}\ga_{n,\ell}^{(1)}\right)\,.
\end{equation}
The identity \eqref{eq:treederivativeformula} is also known as the ``derivative relation'' in the literature \cite{Heemskerk:2009pn,Fitzpatrick:2011dm}: here it descends immediately from the double poles appearing in the spectral integral \eqref{eq:spectralG1}, which tie $a_{n,\ell}^{(1)}$ to $\gamma_{n,\ell}^{(1)}$ as a function of $n$. Notice that operators with $\ell>0$ do not get corrected at this order due to angular momentum conservation. For later convenience, let us specialize to the case of $d=2$,
\begin{equation}\label{eq:gamma1}
    \gamma_{n,\ell}^{(1)}\underset{d=2}{=}\frac{\di_{\ell,0}}{8\pi}\frac1{2\Di_\phi+2n-1}\,,
\end{equation}
\begin{equation}
\begin{aligned}
    a_{n,\ell}^{(1)}\underset{d=2}=
\,\frac{\delta_{\ell,0}\, a^{(0)}_{n,0}}
{8\pi\left(2\Delta_\phi+2n-1\right)}
&\bigg[
2\psi(\Delta_\phi+n)-2\psi(2\Delta_\phi+2n-1)\\
&\quad\left.
+\psi(2\Delta_\phi+n-1)-\psi(n+1)
-\frac{1}{2\Delta_\phi+2n-1}
\right]\,.
\end{aligned}
\end{equation}

In passing, let us comment on the regime $\Delta_\phi\leq d/2$. At $\Delta_\phi = d/2$ the operator $\phi^2$ becomes marginal, and standard and alternative quantization undergo merger-annihilation \cite{Klebanov:1999tb,Kaplan:2009kr,Gorbenko:2018ncu,Lauria:2023uca}. There is a single conformal boundary condition at this value: the spectral function in \eqref{eq:spectralrepprop} has a double pole at $\underline{\Delta} = d/2$, which leads to a possible propagator with logarithmic behavior at the boundary that must be discarded. For $\Delta_\phi < d/2$ (alternative quantization), the contour in \eqref{eq:spectralrepprop} is deformed in order to pick the residue of the pole to the left of the principal series rather than of the pole to the right of the principal series \cite{Ciccone:2024guw}. Furthermore, at $\Delta_\phi = d/4$ it is now the operator $\phi^4$ that becomes marginal, suggesting that the MFT fixed point undergoes merger-annihilation with a nearby interacting fixed point, and the boundary theory becomes unstable for $\Delta_\phi < d/4$ due to the now-relevant $\phi^4$ deformation. 
We also need $\Delta_\phi\geq d/2-1$ to be above the unitarity bound.

\section{One-loop corrections in AdS$_3$} \label{sec:oneloop}
Let us now specialize to $d=2$, namely to $\mathrm{AdS}_3$. In this case the bulk $\lambda\Phi^4$ theory is super-renormalizable. At the level of connected correlators, the only divergent counterterm that is required is the mass counterterm, receiving contributions from the tadpole and sunset diagrams.
Nevertheless, we also have to specify finite normalization conventions to fully determine the renormalization scheme. We fix
the physical mass and the normalization of the two-point function by imposing
\begin{equation}
    \langle \phi(P_1)\phi(P_2)\rangle
    \equiv 
    \frac{\calC_{\Delta_\phi}}{(-2P_1\cdot P_2)^{\Delta_\phi}}\, ,
\end{equation}
with $\Delta_\phi$ now being the physical scaling dimension of $\phi$, related to the physical mass of the bulk field $\Phi$ in the usual way. The remaining finite ambiguity, associated with the definition of
the renormalized quartic coupling, affects the four-point function at one loop by the
addition of a bulk $\Phi^4$ contact diagram. We postpone a discussion of how to get rid of this ambiguity until section \ref{sec:schemeindependent}.

The one-loop correction $\calA^{(2)}$ to the boundary four-point function corresponds to a sum of bulk bubble exchanges in the various channels,
\begin{equation}
    \calA^{(2)}(P_i)=\calA^{(2,s)}(P_i)+\calA^{(2,t)}(P_i)+\calA^{(2,u)}(P_i)\,.
\end{equation}
Our goal is to extract the correction to OPE data by matching with the conformal block expansion
\begin{align}
    \calG^{(2)}(z,\bar{z})&=\sum_{\substack{n\\\ell \text{ even}}}\left(a_{n,\ell}^{(2)}+\frac12\left(a_{n,\ell}^{(0)}\ga_{n,\ell}^{(2)}+a_{n,\ell}^{(1)}\ga_{n,\ell}^{(1)}\right)\frac\de{\de n}+\frac18a_{n,\ell}^{(0)}\left(\ga_{n,\ell}^{(1)}\right)^2\frac{\de^2}{\de n^2}\right)\calK_{\Delta_{n,\ell}^{(0)},\ell}(z,\bar{z})\,.
\end{align}
Note that no additional operators are exchanged at one-loop order.
In what follows, we compute each contribution diagram by diagram.
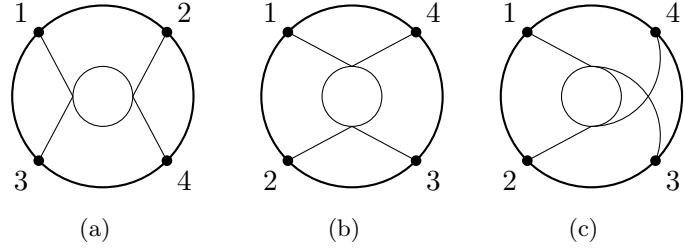
\begin{figure}
\centering\begin{subfigure}[c]{0.15\textwidth}\centering
\begin{tikzpicture}[scale=1]
\def\R{1.2}
    \draw[thick] (0,0) circle (\R);

    \coordinate (p1) at ({-\R*cos(45)}, {\R*sin(45)});
    \coordinate (p4) at ({\R*cos(45)}, {\R*sin(45)});
    \coordinate (p2) at ({-\R*cos(45)}, {-\R*sin(45)});
    \coordinate (p3) at ({\R*cos(45)}, {-\R*sin(45)});
    \fill (p1) circle (2pt);
    \fill (p2) circle (2pt);
    \fill (p3) circle (2pt);
    \fill (p4) circle (2pt);
    \node [anchor=south east] at (p1) {$1$};
    \node [anchor=south west] at (p4) {$2$};
    \node [anchor=north east] at (p2) {$3$};
    \node [anchor=north west] at (p3) {$4$};
    \coordinate (x) at ({-0.33*\R},0);
    \coordinate (y) at ({0.33*\R}, 0);
    
    \draw (p1) -- (x);
    \draw (p2) -- (x);
    \draw (p3) -- (y);
    \draw (p4) -- (y);
    
    \draw (0,0) circle ({0.33*\R});

\end{tikzpicture}
        
        \caption{}
        \end{subfigure}  \qquad
\begin{subfigure}[c]{0.15\textwidth}\centering
\begin{tikzpicture}[scale=1]
\def\R{1.2}
    \draw[thick] (0,0) circle (\R);

    \coordinate (p1) at ({-\R*cos(45)}, {\R*sin(45)});
    \coordinate (p4) at ({\R*cos(45)}, {\R*sin(45)});
    \coordinate (p2) at ({-\R*cos(45)}, {-\R*sin(45)});
    \coordinate (p3) at ({\R*cos(45)}, {-\R*sin(45)});
    \fill (p1) circle (2pt);
    \fill (p2) circle (2pt);
    \fill (p3) circle (2pt);
    \fill (p4) circle (2pt);
    \node [anchor=south east] at (p1) {$1$};
    \node [anchor=south west] at (p4) {$4$};
    \node [anchor=north east] at (p2) {$2$};
    \node [anchor=north west] at (p3) {$3$};

    \coordinate (x) at (0,{0.33*\R});
    \coordinate (y) at (0,{-0.33*\R});

    \draw (p1) -- (x);
    \draw (p4) -- (x);
    \draw (p3) -- (y);
    \draw (p2) -- (y);
    
    \draw (0,0) circle ({0.33*\R});

\end{tikzpicture}
        
        \caption{}
        \end{subfigure}\qquad
\begin{subfigure}[c]{0.15\textwidth}\centering
\begin{tikzpicture}[scale=1]
\def\R{1.2}
    \draw[thick] (0,0) circle (\R);

    \coordinate (p1) at ({-\R*cos(45)}, {\R*sin(45)});
    \coordinate (p4) at ({\R*cos(45)}, {\R*sin(45)});
    \coordinate (p2) at ({-\R*cos(45)}, {-\R*sin(45)});
    \coordinate (p3) at ({\R*cos(45)}, {-\R*sin(45)});
    \fill (p1) circle (2pt);
    \fill (p2) circle (2pt);
    \fill (p3) circle (2pt);
    \fill (p4) circle (2pt);
    \node [anchor=south east] at (p1) {$1$};
    \node [anchor=south west] at (p4) {$4$};
    \node [anchor=north east] at (p2) {$2$};
    \node [anchor=north west] at (p3) {$3$};
    
    \coordinate (x) at (0,{0.33*\R});
    \coordinate (y) at (0,{-0.33*\R});

    \draw (p1) -- (x);

    \draw (p2) -- (y);
    \draw (x) arc (90:-21.5:{0.76*\R});
    \draw (y) arc (-90:21.5:{0.76*\R});
    \draw (0,0) circle ({0.33*\R});

\end{tikzpicture}
        
        \caption{}
        \end{subfigure}
        \caption{One-loop $s$-channel (a), $t$-channel (b), and $u$-channel (c) bubble exchange Witten diagrams.}
        \label{fig:exchange}
        \end{figure}

\subsection{$s$-channel bubble exchange}\label{subsec:sbubble}
To compute the $s$-channel contribution to the one-loop conformal data of $[\phi\phi]_{n,\ell}$, we start from the Witten diagram
\begin{equation}
    \calA^{(2,s)}(P_i)=\frac12\int dX\, dY\, \left(G_{\Delta_\phi}(X,Y)\right)^2 K_{\Delta_\phi}(P_1,X) K_{\Delta_\phi}(P_2,X) K_{\Delta_\phi}(P_3,Y) K_{\Delta_\phi}(P_4,Y)\,,
\end{equation}
and use the spectral representation to write $G^2_{\Di_\phi}(X,Y)$ \cite{Carmi:2018qzm,Carmi:2019ocp},
\begin{equation}
    \left(G_{\Delta_\phi}(X,Y)\right)^2=-i\int d\uDi\, \widetilde{B}_{\Delta_\phi}(\uDi) \Omega_{\uDi}(X,Y)\,,
\end{equation}
with $\widetilde{B}_{\Di_\phi}(\uDi)$ being the spectral function for a bubble of scalars of dimension $\Delta_\phi$ in AdS$_3$,
\begin{equation}
    \widetilde{B}_{\Di_\phi}(\uDi)=-\frac{1}{8\pi(\uDi-1)}\left[\psi\left(\Di_\phi -\frac{\uDi}2\right)-\psi\left(\Di_\phi-\frac{\widetilde\uDi}2\right)\right]\,.
\end{equation}
Proceeding as for the contact diagram, one finds
\begin{equation}
\calG^{(2,s)}(z,\bar z)=-\frac12\int \frac{d\uDi}{2\pi i}\widetilde{B}_{\Di_\phi}(\uDi)\frac{(\uDi-\tuDi)^2}{\calC_{\Di_\phi}^2}b_{\Di_\phi\Di_\phi\uDi}b_{\tuDi \Di_\phi\Di_\phi}K_{\tuDi,0}\calK_{\uDi,0}(z,\bar z)\,.
\end{equation}
We find that the integrand has third-order poles at $\uDi= 2\Delta_\phi+2n$, $n\in\ZZ_{\geq0}$, the additional order coming from the spectral function of the bubble. By closing the contour to the right and picking the appropriate order of derivative of $\calK$, we can read off $\ga_{n,\ell}^{(2,s)}$ and $a_{n,\ell}^{(2,s)}$.  
We can also validate $\ga_{n,\ell}^{(1)}$ and $a_{n,\ell}^{(1)}$ against the tree-level computation. Moreover, the fact that the coefficient of $\de_n^2\calK$ reproduces the known tree-level result $\gamma_{n,0}^{(1)}$ means that in the $t$- and $u$-channel we should find at most double poles at double-trace locations. Explicitly, for the $s$-channel contribution to the anomalous dimensions one finds 
\begin{align}\label{eq:gamma2s}
   \ga_{n,\ell}^{(2,s)}&=\frac{\di_{\ell,0}}{128\pi^2}\frac1{(2\Delta_\phi+2n-1)^2}\left[
\psi(n+1)-\psi(n+2\Delta_\phi-1)
-\frac{2}{2\Delta_\phi+2n-1}
\right]\,.
\end{align}
Moreover, one readily verifies that the following relation holds for the correction to the OPE coefficients,
\begin{equation}\label{eq:oneloopderivatives}
    a_{n,\ell}^{(2,s)}=\frac12\frac{\de}{\de n}\left(a_{n,\ell}^{(0)}\ga_{n,\ell}^{(2,s)}+a_{n,\ell}^{(1)}\ga_{n,\ell}^{(1)}\right)-\frac18 \frac{\de^2}{\de n^2}\left(a_{n,\ell}^{(0)}\left(\ga_{n,\ell}^{(1)}\right)^2\right)\,.
\end{equation}
This is the one-loop generalization of the tree-level derivative relation \eqref{eq:treederivativeformula}, specialized to the $s$-channel contribution. Explicitly, 
\begin{equation}
\begin{aligned}
a_{n,\ell}^{(2,s)}
=
\frac{\di_{\ell,0}}{128\pi^2}&\frac{a_{n,\ell}^{(0)}}{(2n+2\Delta_\phi-1)^2}
\Bigl[
\,2\bigl(\psi(n+\Delta_\phi)-\psi(2n+2\Delta_\phi)\bigr)
\\
&\times
\Bigl(
2\psi(n+\Delta_\phi)
-2\psi(2n+2\Delta_\phi)
+\psi(n+2\Delta_\phi-1)-\psi(n+1)
\Bigr)
\\
&+\psi'(n+\Delta_\phi)
-2\psi'(2n+2\Delta_\phi)
\Bigr]\,.
\end{aligned}
\end{equation}
For more details about the derivative relation \eqref{eq:oneloopderivatives}, we refer to section \ref{subsec:derivativerel}.

\subsection{$t$- and $u$-channel bubble exchange}\label{subsec:tbubble}
Let us focus on the $t$-channel, as the $u$-channel will be entirely analogous to this case. The starting point is the Witten diagram associated to the $t$-channel bubble exchange,
\begin{equation}
\begin{aligned}
    \calA^{(2,t)}(P_i)&=\frac1{2}\int dX\, dY\, \left(G_{\Delta_\phi}(X,Y)\right)^2K_{\Delta_\phi}(P_1,X) K_{\Delta_\phi}(P_2,Y) K_{\Delta_\phi}(P_3,Y) K_{\Delta_\phi}(P_4,X)\\
        &=-\frac14\int \frac{d\uDi}{2\pi i}\widetilde{B}_{\Di_\phi}(\uDi)(\uDi-\tuDi)^2b_{\Di_\phi\Di_\phi\uDi}b_{\tuDi \Di_\phi\Di_\phi}\Psi_{\uDi,0}^t(P_i)\,,
\end{aligned}
\end{equation}
where $\Psi_{\uDi,0}^t(P_1,P_2,P_3,P_4)=\Psi_{\uDi,0}^s(P_1,P_4,P_2,P_3)$ is the $t$-channel scalar conformal partial wave of dimension $\uDi$. The latter can in turn be expressed in terms of $s$-channel partial waves by means of the conformal $6j$ symbol \cite{Liu:2018jhs,Meltzer:2019nbs},
\begin{equation}
    \Psi_{\uDi,0}^{t}(P_i)=\sum_{J'=0}^\infty\int_+ \frac{d\uDi'}{2\pi i}\left\{\begin{array}{ccc}
        \phi & \phi & [\uDi,0] \\
        \phi & \phi & [\uDi',J']
    \end{array}\right\}\frac1{n_{\uDi',J'}}\Psi_{\uDi',J'}^{s}(P_i)\,,
\end{equation}
where $n_{\Delta',J'}$, in $d=2$, is defined as
\begin{equation}
    n_{\Delta, J}=\frac\pi 2 K_{\Delta, J}K_{\widetilde{\Delta}, J}\,.
\end{equation}
and $\int_+ d\uDi'$ means that the integral is over the upper half of the principal series $\frac d2+i\RR_+$. The $6j$ symbol itself can be decomposed in terms of what we will call ``reduced $6j$ symbols'',
\begin{equation}
    \left\{\begin{array}{ccc}
        \phi & \phi & [\uDi,0] \\
        \phi & \phi & [\uDi',J']
    \end{array}\right\}=K_{\tuDi,0}\begin{pmatrix}
        \phi&\phi&[\uDi,0]\\
        \phi&\phi&[\uDi',J']
    \end{pmatrix}+ K_{\uDi,0}\begin{pmatrix}
        \phi&\phi&[\tuDi,0]\\
        \phi&\phi&[\uDi',J']
    \end{pmatrix}\,.
\end{equation}
Then, extending the contour to the whole principal series,
\begin{equation}
\begin{aligned}
    \calG^{(2,t)}(z,\bar z)&=-\int \frac{d\uDi}{2\pi i}\widetilde{B}_{\Di_\phi}(\uDi)\frac{(\uDi-\tuDi)^2}{\calC_{\Di_\phi}^2}b_{\Di_\phi\Di_\phi\uDi}b_{\tuDi \Di_\phi\Di_\phi}K_{\tuDi,0}\\
    &\qquad \times \sum_{J'} \int \frac{d\uDi'}{2\pi i}\left(\begin{array}{ccc}
        \phi & \phi & [\uDi,0] \\
        \phi & \phi & [\uDi',J']
    \end{array}\right)\frac1{\pi}K_{\uDi',J'}\calK_{\uDi',J'}(z,\bar z)\,.
    \end{aligned}
\end{equation}
In our case, that is in $d=2$ and equal external scaling dimensions, the reduced $6j$ symbol takes a particularly symmetric form,
\begin{equation}\label{eq:reduced6j}
    \begin{pmatrix}
        \phi&\phi&[\uDi,0]\\
        \phi&\phi&[\uDi',J']
    \end{pmatrix}=\frac{(-1)^{J'}}{4}\,
\frac{
\Gamma(\underline{h}')^2\Gamma(\widetilde{\bar{\underline{h}}}{}' )^2
}{
\Gamma(2\underline{h}')\Gamma(2\widetilde{\bar{\underline{h}}}{}')
}
\,
\sin^2\!\left(
\pi\left({\uDi}/{2}-\Delta_\phi\right)
\right)\,
\widehat\Omega_\phi(\uDi,\underline{h}')\,
\widehat\Omega_\phi(\uDi,\widetilde{\bar{\underline{h}}}{}')\,,
\end{equation}
where we have introduced 
\begin{equation}
\underline{h}' \equiv \frac{\uDi'+J'}{2}\,,
\qquad
\widetilde{\bar{\underline{h}}}{}' \equiv \frac{\tuDi'+J'}{2}\,,
\end{equation}
and we have let\footnote{The reduction of (3.38) of \cite{Liu:2018jhs} to a very-well-poised ${}_7F_6(1)$ is not special to our case. In their notation, letting $a\equiv h' + 2h - p$, one finds that a possible representation for $\Omega_{h,h',p}^{h_i}$ is
\begin{equation*}
\begin{aligned}
\Omega^{h_i}_{h,h',p}
&=\frac{
\Gamma(h'-p+1)
\Gamma(h'-h_{12}+h_{34}-p+1)
\Gamma(h'+2h-p+1)
\Gamma(h_{13}+h+p-1)
\Gamma(h_{42}+h+p-1)
}{
\Gamma(h'+p-1)
\Gamma(h'-h_{12}+h-p+1)
\Gamma(h'+h_{13}+h)
\Gamma(h'+h_{42}+h)
\Gamma(h'+h_{34}+h-p+1)
}\\
&\ \times 
\Gamma(2h')\,
{}_7F_6\!\left(
\begin{matrix}
a,\;
1+\tfrac{a}{2},\;
1+h-p-h_{13},\;
1+h-p-h_{42},\;
h-h_{34},\;
h'-p+1,\;
h_{12}+h
\\
\tfrac{a}{2},\;
h'+h_{13}+h,\;
h'+h_{42}+h,\;
h'+h_{34}+h-p+1,\;
2h,\;
h'-h_{12}+h-p+1
\end{matrix}
;1
\right)\,.
\end{aligned}
\end{equation*}

} 
\begin{equation}
\begin{aligned}
\widehat\Omega_\phi(\uDi,x)
&=\frac{
\Gamma(\uDi)\,
\Gamma(a+1-3x)\,
\Gamma(2x-b)^2
}{
\Gamma\!\left(\frac{\uDi}{2}\right)^2\,
\Gamma(a+x+1-2b)
}
{}_4F_3\!\left(
\begin{matrix}
x,\;x,\;2x-b,\;2x-b\\ 
2x,\;3x-a,\;a+x+1-2b
\end{matrix}
;1
\right)
\\
&\ + \frac{
\Gamma(2x)\,
\Gamma(3x-a-1)\,
\Gamma(a+1-2x)^2
}{
\Gamma(x)^2\,
\Gamma(a+1-x)
}
{}_4F_3\!\left(
\begin{matrix}
\frac{\uDi}{2},\;\frac{\uDi}{2},\;a+1-2x,\;a+1-2x\\ 
\uDi,\;a+2-3x,\;a+1-x
\end{matrix}
;1
\right)\\
&=
\mathcal P_\phi(\uDi,x)\,
{}_7W_6\!\left(a;\begin{array}{c}
     b,b,x  \\
     x,a+1-2x 
\end{array}\right)\,,
\end{aligned}
\end{equation}
with 
\begin{equation}
a=\frac{\uDi}{2}+2x-\Delta_\phi\,,
\qquad
b=1+x-\Delta_\phi\,,
\end{equation}
and ${}_7W_6$ is a shorthand for a very-well-poised ${}_7F_6(1)$ \cite{Bailey:1935},
\begin{equation}
   {}_7W_6\!\left(a;\begin{array}{c}
     b_1,b_2,b_3  \\
     b_4,b_5 
\end{array}\right) \equiv {}_7F_6\!\left(
\begin{matrix}
a,
1+\tfrac{a}{2},
b_1,
b_2,
b_3,
b_4,
b_5
\\
\tfrac{a}{2},
\mbox{\small $1+a-b_1,
1+a-b_2,
1+a-b_3,
1+a-b_4,
1+a-b_5$}
\end{matrix}
;1
\right)\,,
\end{equation}
and 
\begin{equation}
\mathcal P_\phi(\uDi,x)
=
\frac{
\Gamma(\uDi)\,
\Gamma\!\left(1+\frac{\uDi}{2}-\Delta_\phi\right)^2\,
\Gamma\!\left(1+\frac{\uDi}{2}+2x-\Delta_\phi\right)\,
\Gamma(x+\Delta_\phi-1)^2
}{
\Gamma\!\left(\frac{\uDi}{2}+\Delta_\phi-1\right)\,
\Gamma\!\left(1+x+\frac{\uDi}{2}-\Delta_\phi\right)^2\,
\Gamma\!\left(x+\frac{\uDi}{2}\right)^2
}\,.
\end{equation}

A feature of the reduced $6j$ symbol is that it does not have poles in $\uDi$ to the right of the principal series. We can then swap $\int d\uDi$ with $\sum_{J'}\int d\uDi'$ and close the contour. The $\sin^2(\pi(\uDi/2-\Di_\phi))$ term in the reduced $6j$ symbol kills the scalar double-trace poles in the three-point coefficients $b_{\Delta_\phi\Delta_\phi \uDi}b_{\tuDi \Di_\phi\Di_\phi}$, thus the only relevant poles are the ones of the spectral function $\widetilde{B}_{\Di_\phi}(\uDi)$. They satisfy
\begin{equation}\label{eq:bubbleres}
    \Res_{\uDi=\uDi_p}\widetilde{B}_{\Di_\phi}(\uDi)=-\frac1{4\pi(\uDi_p-1)}\,,\quad \uDi_p=2\Di_\phi+2p\,,\quad p\in \ZZ_{\geq0}\,.
\end{equation}
Therefore,
\begin{equation}
    \begin{aligned}
        \calG^{(2,t)}(z,\bar{z})=-\frac1{4\pi^2\calC_{\Di_\phi}^2}\sum_{J'=0}^\infty\int \frac{d\uDi'}{2\pi i}&\left[\sum_{p=0}^\infty \frac{(\uDi_p-\tuDi_p)^2}{\uDi_p-1}b_{\Di_\phi\Di_\phi\uDi_p}b_{\tuDi_p \Di_\phi\Di_\phi}K_{\tuDi_p,0}\right.\\
        &\qquad\qquad \left. \times \left(\begin{array}{ccc}
        \phi & \phi & [\uDi_p,0] \\
        \phi & \phi & [\uDi',J']
    \end{array}\right)K_{\uDi',J'}\calK_{\uDi',J'}(z,\bar z)\right]\,.
    \end{aligned}
\end{equation}
The reduced $6j$ symbol has double and simple poles at double-trace locations $\uDi'\to\Di_{n,J'}^{(0)}=2\Di_\phi+2n+J'$, $n\in\ZZ_{\geq0}$, 
\begin{equation}
    \widehat\Omega_\phi(\uDi_p,\widetilde{\bar{\underline{h}}}{}')\sim  \frac{\widehat\omega_\phi(\uDi_p)}{(\uDi'-\Di_{n,J'}^{(0)})^2}\,, 
\end{equation}
with
\begin{equation}\label{eq:smallomega}
    \widehat\omega_\phi(\uDi_p)=\frac{4
\Gamma(\uDi_p)\,
\Gamma\!\left(n+\frac{\uDi_p}{2}\right)\,
}{(n!)^2
\Gamma\!\left(\frac{\uDi_p}{2}\right)^2\,
\Gamma\!\left(\frac{\uDi_p}{2}-n\right)
}
{}_4F_3\!\left(
\begin{matrix}
1-\Delta_\phi-n,1-\Delta_\phi-n,-n,-n\\
2-2\Delta_\phi-2n,1-n-\frac{\uDi_p}{2},
\frac{\uDi_p}{2}-n
\end{matrix}
;1
\right)\,.
\end{equation}
Then, closing the contour on the right, we arrive at
\begin{equation}
\begin{aligned}
    \calG^{(2,t)}(z,\bar z)=\frac1{4\pi^2\calC_{\Di_\phi}^2}&\sum_{n=0}^\infty \sum_{\ell=0}^\infty\sum_{p=0}^\infty\frac{(\uDi_p-\tuDi_p)^2}{\uDi_p-1}b_{\Di_\phi\Di_\phi\uDi_p}b_{\tuDi_p \Di_\phi\Di_\phi}K_{\tuDi_p,0}\\
    &\quad\times \Res_{\uDi'=\Di_{n,\ell}^{(0)}}\left[\left(\begin{array}{ccc}
        \phi & \phi & [\uDi_p,0] \\
        \phi & \phi & [\uDi',\ell]
    \end{array}\right)K_{\uDi',\ell}\calK_{\uDi',\ell}(z,\bar z)\right]\,,
    \end{aligned}
\end{equation}
where we renamed $J'\to \ell$. Comparing with the expected expansion, we obtain an infinite-sum representation of the $t$-channel contribution to the one-loop anomalous dimension $\gamma_{n,\ell}^{(2,t)}$,
\begin{equation}\label{eq:infinitegamma2t}
\begin{aligned}
    \gamma_{n,\ell}^{(2,t)}=\frac1{4\pi^2a_{n,\ell}^{(0)}\calC_{\Di_\phi}^2}&\sum_{p=0}^\infty \frac{(\uDi_p-\tuDi_p)^2}{\uDi_p-1}b_{\Di_\phi\Di_\phi\uDi_p}b_{\tuDi_p \Di_\phi\Di_\phi}K_{\tuDi_p,0} \\
    &\quad \times \left.\left[\left(\uDi'-\uDi_{n,\ell}^{(0)}\right)^2\left(\begin{array}{ccc}
        \phi & \phi & [\uDi_p,0] \\
        \phi & \phi & [\Di_{n,\ell}^{(0)},\ell]
    \end{array}\right)K_{\uDi',\ell}\right]\right|_{\uDi'=\Di_{n,\ell}^{(0)}}\,,
\end{aligned}
\end{equation}
and equivalently for the OPE coefficients,
\begin{equation}\label{eq:infinitea2t}
\begin{aligned}
    a_{n,\ell}^{(2,t)}=\frac1{4\pi^2\calC_{\Di_\phi}^2}&\sum_{p=0}^\infty \frac{(\uDi_p-\tuDi_p)^2}{\uDi_p-1}b_{\Di_\phi\Di_\phi\uDi_p}b_{\tuDi_p \Di_\phi\Di_\phi}K_{\tuDi_p,0} \\
    &\quad \times\left.\frac{d}{d\uDi'} \left[\left(\uDi'-\uDi_{n,\ell}^{(0)}\right)^2\left(\begin{array}{ccc}
        \phi & \phi & [\uDi_p,0] \\
        \phi & \phi & [\Di_{n,\ell}^{(0)},\ell]
    \end{array}\right)K_{\uDi',\ell}\right]\right|_{\uDi'=\Di_{n,\ell}^{(0)}}\,.
\end{aligned}
\end{equation}
Notice that the correction to the OPE coefficients computed in this way automatically satisfies the derivative relation
\begin{equation}\label{eq:oneloopderivativet}
    a_{n,\ell}^{(2,t)}=\frac12\frac\de{\de n}\left(a_{n,\ell}^{(0)}\gamma_{n,\ell}^{(2,t)}\right)\,, 
\end{equation}
provided that the analytic $n$-dependence of $\ga_{n,\ell}^{(2,t)}$ is fixed as in \eqref{eq:infinitegamma2t}. For more comments on this relation, we refer to section \ref{subsec:derivativerel}.

In Appendix \ref{sec:proof} we prove that, for generic $\Delta_\phi>1$, the anomalous dimension $\gamma_{n,\ell}^{(2,t)}$ can be written as a finite sum over very-well-poised ${}_7F_6(1)$ functions. A possible representation is
\begin{equation}\label{eq:finitegamma2t}
    \begin{aligned}
        \gamma_{n,\ell}^{(2,t)}=\frac{(-1)^{\ell+1}}{128\pi^2} \sum_{q=0}^n c_{n,q} \Lambda_{m,q} 
         { }_7 W_6\left(a_q;\begin{array}{c}2 \Di_\phi-1,\Di_\phi+q,\Di_\phi+q\\
         m+2 \Di_\phi+q-\frac{1}{2},1\end{array}\right)\,,
    \end{aligned}
\end{equation}
where $a_q=m+3\Di_\phi+q-1$, $m=n+\ell$, and
\begin{equation}\label{eq:cnq}
    c_{n,q}=\frac{\left(\Delta_\phi-\frac12\right)_n}
{\left(\Delta_\phi+\frac12\right)_n}
\,
\frac{
(-n)_q
\left(n+2\Delta_\phi-1\right)_q
\left(\Delta_\phi-\frac12\right)_q^2
\left(\frac{\Delta_\phi}{2}+\frac34\right)_q
}{
q!\,(1)_q
\left(\frac32-\Delta_\phi-n\right)_q
\left(n+\Delta_\phi+\frac12\right)_q
\left(\frac{\Delta_\phi}{2}-\frac14\right)_q
}\,.
\end{equation}
\begin{equation}\label{eq:Lambdamq}
    \Lambda_{m,q}=\frac{\Gamma(2 \Di_\phi-1) \Gamma(m+1)(m+3 \Di_\phi+q-1)}{(m+\Di_\phi+q) \Gamma(m+2 \Di_\phi)(2 \Di_\phi-1)(m+2 \Di_\phi-1)}\,.
\end{equation}
The truncation of the $q$-sum in \eqref{eq:finitegamma2t} at $n$ comes from the $(-n)_q$ term in $c_{n,q}$. 

The $u$-channel treatment is analogous and leads to formally identical results up to an additional $(-1)^\ell$ factor.

\section{Some specialized results}\label{sec:specialized}
The closed-form expression \eqref{eq:finitegamma2t} is valid for arbitrary $n,\ell$, and $\Delta_\phi > 1$. In this section we record several limits and specializations that either yield simpler expressions or connect directly to prior results in the literature.

\subsection{Special values of $\Delta_\phi$} \label{subsec:speecialdelta}
While \eqref{eq:finitegamma2t} is a closed-form expression, the trajectory label $n$ also appears as the upper limit of a sum. In some cases, namely when $\Delta_\phi$ is a half-integer, we are able to reformulate the expressions in such a way that the termination depends on the value of $\Delta_\phi$ instead.

Let $\Delta_\phi=3/2+\alpha$, $\alpha\in \ZZ_{\geq0}$. The very-well-poised $_7F_6(1)$ reduces to a terminating $_4F_3(1)$,
$${}_7W_6\left(a_q;\begin{array}{c}2 \Di_\phi-1,\Di_\phi+q,\Di_\phi+q\\
         m+2 \Di_\phi+q-\frac{1}{2},1\end{array}\right)=P_{m,q}\,{}_4F_3\left(\begin{matrix}
    -\al,q+\al+\tfrac32,q+\al+\tfrac32,1\\
    \al+2,m+q+\al+\tfrac32,q-m-\al+\tfrac12
\end{matrix};1\right)\,,$$
with $P_{m,q}$ such that
$$\Lambda_{m,q}P_{m,q}=\frac{\Ga(2\al+2)\Ga(m+1)}{(2\al+2)\Ga(m+2\al+2)}\frac{1}{(m+\al+q+\tfrac32)(m+\al-q+\tfrac12)}\,.$$
Moreover, one has
$$c_{n,q}=\frac{(\al+1)_n}{(\al+2)_n} D_q\,,\qquad D_q=\frac{(-n)_q(n+2\al+2)_q(\al+1)_q^2(\tfrac{\al+3}2)_q}{q!(1)_q(-n-\al)_q(n+\al+2)_q(\tfrac{\al+1}2)_q}\,.$$
Now, we notice that $F_\al(q)\equiv{}_4F_3(1)/((m+\al+q+\tfrac32)(m+\al-q+\tfrac12))$ admits the following partial fraction decomposition, 
$$F_\al(q)=\sum_{s=0}^\al C_s\left[\frac1{q+m+s+\al+\tfrac32}-\frac1{q-m-s-\tfrac12}\right]\,,$$
with
$$C_s=\frac{(m+\al+2)_\al^2}{(\al+2)_\al(2m+\al+2)_{\al+1}}\frac{(-\al)_s(m+1)_s^2(2m+\al+2)_s}{s!(m+\al+2)_s^2(2m+2\al+3)_s}\,.$$
Now we perform the $q$ sum. To do so, let $H(z)\equiv\sum_{q=0}^n D_q/(q+z)$. One finds
$$\begin{aligned}
    H(z)-H(\al+1-z)&=\frac{\al+1-2z}{z(\al+1-z)}{}_7W_6\left(\al+1;\begin{matrix}
    -n,n+2\al+2,\al+1\\ z,\al+1-z
\end{matrix}\right)\\
&=\frac{\al+1-2z}{z(\al+1-z)}\frac{(\al+2)_n(2\al+2)_n}{(\al+1)_n n!}{}_4F_3\left(\begin{matrix}
    -n,1,\al+1,n+2\al+2\\2\al+2,1+z,\al+2-z
\end{matrix};1\right)\,,
\end{aligned}$$
which we can apply to $z=m+s+\al+\tfrac32$ in our case. The final result is
\begin{equation}
\begin{aligned}
    &\ga_{n,\ell}^{(2,t)}\left(\al+\tfrac32\right)=\frac{(-1)^{\ell+1}}{128\pi^2}\frac{\Ga(\al+1)(2\al+2)_n(\al+2)_\al}{\Ga(n+1)(m+1)_{\al+1}(m+\al+2)_\al}\\
    &\!\!\!\!\!\!\!\times \sum_{s=0}^\al C_s\left(\frac{1}{2m+2s+1}+\frac1{2m+2\al+2s+3}\right){}_4F_3\left(\begin{matrix}
    -n,1,\al+1,n+2\al+2\\2\al+2,m+s+\al+\tfrac52,\tfrac12-m-s
\end{matrix};1\right)\,.
\end{aligned}
\end{equation}
In particular, for $\al=0$, that is for $\Delta_\phi=3/2$, the expression further collapses to
\begin{equation}
    \gamma_{n,\ell}^{(2,t)}\left(\frac32\right)=\frac{(-1)^{\ell+1}}{256\pi^2}\frac{2\psi\left(\ell+n+\tfrac32\right)-\psi\left(\ell+2n+\tfrac52\right)-\psi\left(\ell+\tfrac12\right)}{(n+1)(\ell+n+1)}\,,
\end{equation}
which matches the results of \cite{Xiao:2026prw} for $n=0$. In general, we find that similar decompositions in terms of digamma functions evaluated at half-integers exist for any value of $\al$, typically with rational correction terms. This form can be obtained by explicitly performing the polynomial division $D_q/(q+z)$.

We were not able to find similar reductions for other families of $\Delta_\phi$, such as integer $\Delta_\phi$. However, for $\Delta_\phi=2$ we were able to observe the following,
\begin{equation}
    \gamma_{0,\ell}^{(2,t)}(2)=\frac{(-1)^{\ell+1}}{128\pi^2} \frac{2(\ell+1)(\ell+2)}{(2\ell+1)(2\ell+3)(2\ell+5)}\left[\psi'(\ell+1)-\frac{(2\ell+1)(\ell^2+5\ell+5)}{2(\ell+1)^2(\ell+2)^2}\right]\,,
\end{equation}
which for low values of $\ell$ reduces to the results of \cite{Aharony:2016dwx}.

\subsection{Large-spin limit} \label{subsec:largespin}
The large-spin regime, $\ell \gg 1$ at fixed $n$, is the domain of the lightcone bootstrap. In this limit, the closed-form expression \eqref{eq:finitegamma2t} admits a systematic expansion in inverse powers of the conformal spin, whose structure is predicted on general grounds by large-spin perturbation theory \cite{Fitzpatrick:2012yx,Komargodski:2012ek,Alday:2016njk,Caron-Huot:2017vep}.

\paragraph{Leading Regge trajectory.} For $n=0$, the sum in \eqref{eq:finitegamma2t} reduces to a single term. It is then interesting to perform an expansion at large spin $\ell$, or more conveniently at large conformal spin, to compare with  lightcone bootstrap predictions. We define the conformal spin as ${\sf J}_n^2=(\tau_n/2+\ell)(\tau_n/2+\ell-1)$, where $\tau_n=2\Di_\phi+2n$ is the twist, and we denote ${\sf J}\equiv {\sf J}_0$, $\tau\equiv \tau_0$. One finds that the large $\sf J$ expansion takes the form
\begin{equation}
\gamma^{(2,t)}_{0,\ell}(\Delta_\phi)
\underset{{\sf J}\to \infty}\sim
\frac{(-1)^{\ell+1}}{128\pi^2}\,
\frac{\Gamma(\tau-1)}{\tau-1}\,
{\sf J}^{-\tau} \left[1+ \sum_{k=1}^{\infty}\frac{\left(\frac\tau 2\right)_k}{\left(\frac{\tau+1}2\right)_k} \frac{P_k(\tau)}{{\sf J}^{2k}}\right]\,,
\end{equation}
where $P_k(\tau)$ are polynomials with rational coefficients. This matches the lightcone bootstrap expectation that only even powers of $\sf J$ enter in the subleading corrections \cite{Alday:2015eya,Kaviraj:2015cxa,Kaviraj:2015xsa}. For instance, the first terms read
\begin{align}
P_1(\tau)&=\frac{\tau^3+\tau^2-16\tau+8}{24}\,,\\
P_2(\tau)&=\frac{5\tau^6+28\tau^5-129\tau^4-928\tau^3
+3616\tau^2-2520\tau+288}{5760}\,,\\
P_3(\tau)&=\mbox{\small 
$\frac{35\tau^9+483\tau^8+429\tau^7-23787\tau^6
-53136\tau^5+812592\tau^4
-1881464\tau^3+1162272\tau^2+6336\tau-69120}{
2903040
}$}
\,.
\end{align}
Let us now look at some special cases. For $\Delta_\phi=3/2$, namely $\tau=3$, this gives
\begin{equation}
\gamma^{(2,t)}_{0,\ell}\left(\frac 32\right)
\underset{{\sf J}\to \infty}\sim
\frac{(-1)^{\ell+1}}{256\pi^2}\,
{\sf J}^{-3}
\left[
1-\frac{1}{8{\sf J}^2}
+\frac{3}{128{\sf J}^4}
-\frac{5}{1024{\sf J}^6}
+\dots
\right]\,,
\end{equation}
in accordance with \cite{Xiao:2026prw}.
For $\Delta_\phi=2$, namely $\tau=4$, this gives
\begin{equation}
\gamma^{(2,t)}_{0,\ell}(2)
\underset{{\sf J}\to \infty}\sim
\frac{(-1)^{\ell+1}}{192\pi^2}\,
{\sf J}^{-4}
\left[
1+\frac{4}{5{\sf J}^2}
+\frac{4}{7{\sf J}^4}
+\frac{16}{35{\sf J}^6}
+\dots
\right]\,,
\end{equation}
which matches the analysis of \cite{Aharony:2016dwx}.

\paragraph{Subleading trajectories.} We can repeat the large ${\sf J}_n$ expansion for subleading trajectories $n>0$, $n$ fixed. Here we report only the expression containing the first correction in terms of the minimal twist $\tau=2\Delta_\phi$,
\begin{equation}
\gamma^{(2,t)}_{n,\ell}(\Delta_\phi)
\underset{{\sf J_n}\to \infty}\sim
\frac{(-1)^{\ell+1}}{128\pi^2}\,
\frac{\Gamma(\tau-1+n)}{n!(\tau-1)}\,
{\sf J}_n^{-\tau} \left[1+\tfrac{\tau P_1(\tau) +n(\tau-1)(\tau+n-1)}{\tau+1}{\sf J}_n^{-2}+\dots \right]\,.
\end{equation}

\subsection{High-energy limits}\label{subsec:highenergy}
We consider two complementary high-energy limits: the Regge limit and the bulk-point limit. Both correspond to high-energy kinematic regimes of the boundary four-point function, associated respectively to fixed AdS impact parameter \cite{Cornalba:2006xm, Cornalba:2007zb, Costa:2012cb} and to the approach to the bulk-point singularity \cite{Gary:2009ae,Maldacena:2015iua}. For concreteness, we work with $\Delta_\phi = 3/2$ throughout.

\paragraph{Regge limit.} We consider the so-called Regge limit,
$$n,\ell\to \infty\,,\qquad \ell/n \text{ fixed}\,. $$
This is equivalent to considering a fixed AdS impact parameter $b\approx\log(\ell/n+1)$. One finds
\begin{equation}
    \ga_{n,\ell}^{(2,t)}\left(\frac 32\right)\underset{\rm Regge}{\approx}\frac{(-1)^{\ell+1}}{128\pi^2}\frac{1}{2n^2 e^b}\left(-\log\left(1-e^{-2b}\right)\right)\,.
\end{equation}
Several comments are in order. We recognize that this decays at large $n$ as $n^{-2}$, which is the same power observed in tree-level $t$-channel scalar exchanges: in particular, at one loop no enhancement occurs. On the other hand, the dependence on $e^b$ includes a logarithm. Expanding the logarithm,
$$-\frac1{2e^{b}}\log\left(1-e^{-2b}\right)=\sum_{k=0}^\infty \frac{e^{-(2k+3)b}}{2k+2}\,,$$
we recognize that each term corresponds to the exchange of the operator $[\phi\phi]_{k,0}$, of dimension $\Delta_{k,0}=2k+3$, and the weight $1/(2k+2)$ is related to the residue of the bubble at the corresponding double-trace pole, cf. \eqref{eq:bubbleres}. At large impact parameter, the $k=0$ term dominates: this is consistent with the expectation that in the large $b$ limit scattering is controlled by the lightest exchanged state. The limit of small $b$, instead, leads to a $\log(b)$ divergence that signals the breakdown of the Regge expansion, since it corresponds to $\ell\ll n$. Physics in this limit is better described in the bulk-point limit.

\paragraph{Bulk-point limit.} We now consider the bulk-point limit,
$$n\to \infty\,,\qquad \ell \text{ fixed}\,.$$ 
The leading large-$n$ behavior is
\begin{equation}
    \ga_{n,\ell}^{(2,t)}\left(\frac 32\right)\underset{\rm b.p.}{\approx}\frac{(-1)^{\ell+1}}{128\pi^2} \frac{\log(n)}{2n^2}\,.
\end{equation}
The physical origin of the $\log(n)$ term is the same as in the Regge limit. In that case, the whole tower of double-trace exchanges $[\phi\phi]_{k,0}$ contributed to the anomalous dimension at fixed impact parameter $b$. In this case, only the lightest $n$ states are kinematically accessible, and thus, for $n$ large,
$$
    \frac12\log (n) \approx \sum_{k=0}^{n} \frac{1}{2k+2}\,.
$$

\section{Discussion}\label{sec:discussion}
In this paper, we have obtained closed-form expressions for the one-loop corrections to the CFT data of all double-trace operators $[\phi\phi]_{n,\ell}$, valid for arbitrary $n$, $\ell$, and $\Delta_\phi > 1$, for the boundary theory corresponding to $\lambda\Phi^4$ in AdS$_3$. Explicitly, for the anomalous dimensions,
\begin{equation}
    \gamma_{n,\ell}^{(2)}=\gamma_{n,0}^{(2,s)}\di_{\ell,0} + \big(1+(-1)^\ell\big)\gamma_{n,\ell}^{(2,t)}\,,
\end{equation}
with 
$$\ga_{n,0}^{(2,s)}=\frac{1}{128\pi^2}\frac1{(2\Delta_\phi+2n-1)^2}\left[
\psi(n+1)-\psi(n+2\Delta_\phi-1)
-\frac{2}{2\Delta_\phi+2n-1}
\right]\,,$$
and
\begin{equation*}
    \begin{aligned}
        \gamma_{n,\ell}^{(2,t)}=\frac{(-1)^{\ell+1}}{128\pi^2} \sum_{q=0}^n c_{n,q} \Lambda_{m,q} 
         { }_7 W_6\left(a_q;\begin{array}{c}2 \Di_\phi-1,\Di_\phi+q,\Di_\phi+q\\
         m+2 \Di_\phi+q-\frac{1}{2},1\end{array}\right)\,,
    \end{aligned}
\end{equation*}
where $a_q=m+3\Di_\phi+q-1$, $m=n+\ell$, and $c_{n,q}$, $\Lambda_{m,q}$ are given by \eqref{eq:cnq}, \eqref{eq:Lambdamq}, respectively. Correspondingly, the OPE coefficients are expressed in terms of derivative relations involving the anomalous dimensions.

Beyond the explicit results and their specialization in various limits, we collect here a number of remarks on their structure and properties.

\subsection{Scheme ambiguity and scheme-independent parametrization}\label{sec:schemeindependent}
Even in a super-renormalizable theory like $\lambda\Phi^4$ in three spacetime dimensions, subleading anomalous dimensions are scheme-dependent and thus ambiguous unless a scheme is fixed. In the minimal
theory this ambiguity is generated by a finite redefinition of the quartic coupling, $\lambda\to \lambda\, (1+c_\la \,\la + \dots)$, or equivalently by adding a finite local $\Phi^4$ contact diagram.
At one loop, then, it shifts only the spin-zero anomalous dimensions and corrections to OPE coefficients,
\begin{equation}
    \gamma_{n,\ell}^{(2)}
    \sim
    \gamma_{n,\ell}^{(2)}
    +
    c_\lambda\,\gamma_{n,0}^{(1)}\delta_{\ell,0}\,,\qquad a_{n,\ell}^{(2)}\sim a_{n,\ell}^{(2)}+c_\la\, a_{n,0}^{(1)}\di_{\ell,0}\,.
\end{equation}

A convenient way to remove this ambiguity is not to define observables in terms
of the bulk coupling  $\lambda $ and fix a renormalization scheme, but rather in terms of a physical CFT
datum. This is analogous in spirit to an on-shell scheme in flat-space QFT, where the renormalized parameters are defined in terms of physical observables.  We choose as perturbative parameter the full anomalous dimension of the
lowest spin-zero double-trace operator,
\begin{equation}
    g \equiv \gamma_{0,0}=\Delta_{0,0}-2\Delta_\phi\,.
\end{equation}
In an arbitrary intermediate scheme we can invert the perturbative expansion. Up to one loop, one has
\begin{equation}
    \gamma_{0,0}
    =
    \lambda\,\gamma_{0,0}^{(1)}
    +
    \lambda^2\,\gamma_{0,0}^{(2)}
    +
    O(\lambda^3) \implies 
    \lambda
    =
    \frac{g}{\gamma_{0,0}^{(1)}}
    -
    \frac{\gamma_{0,0}^{(2)}}{\big(\gamma_{0,0}^{(1)}\big)^3}
    g^2
    +
    O(g^3)\,.
\end{equation}
Substituting this relation into the remaining CFT data gives a physical
expansion in  $g $. For example,
\begin{equation}
    \gamma_{n,\ell}
    =
    g\,\widehat\gamma_{n,\ell}^{(1)}
    +
    g^2\,\widehat\gamma_{n,\ell}^{(2)}
    +
    O(g^3)\,,
\end{equation}
with
\begin{equation}\label{eq:gammahat}
    \widehat\gamma_{n,\ell}^{(1)}
    \equiv
    \frac{\gamma_{n,\ell}^{(1)}}{\gamma_{0,0}^{(1)}},
    \qquad
    \widehat\gamma_{n,\ell}^{(2)}
    \equiv
    \frac{\gamma_{n,\ell}^{(2)}}{\big(\gamma_{0,0}^{(1)}\big)^2}
    -
    \frac{\gamma_{0,0}^{(2)}\gamma_{n,\ell}^{(1)}}
    {\big(\gamma_{0,0}^{(1)}\big)^3}\,.
\end{equation}
Similarly, one obtains
\begin{equation}
    a_{n,\ell}
    =
    a_{n,\ell}^{(0)}
    +
    g\,\widehat a_{n,\ell}^{(1)}
    +
    g^2\,\widehat a_{n,\ell}^{(2)}
    +
    O(g^3)\,,
\end{equation}
where
\begin{equation}\label{eq:ahat}
    \widehat a_{n,\ell}^{(1)}
    =
    \frac{a_{n,\ell}^{(1)}}{\gamma_{0,0}^{(1)}}\,,
    \qquad
    \widehat a_{n,\ell}^{(2)}
    =
    \frac{a_{n,\ell}^{(2)}}{\big(\gamma_{0,0}^{(1)}\big)^2}
    -
    \frac{\gamma_{0,0}^{(2)}a_{n,\ell}^{(1)}}
    {\big(\gamma_{0,0}^{(1)}\big)^3}\,.
\end{equation}
The hatted coefficients are invariant under finite redefinitions of the coupling $\lambda$, because the induced shift of  $\gamma_{n,\ell}^{(2)} $ is exactly canceled by the corresponding shift of  $\gamma_{0,0}^{(2)} $. Thus, after fixing the two-point normalization of $\phi$, expressing all CFT data in terms of  $g=\gamma_{0,0} $ gives a scheme-independent parametrization of the perturbative answer.

\subsection{Derivative relations} \label{subsec:derivativerel}
For the case of $\lambda\Phi^4$ theory in AdS$_3$, we can analyze how the derivative relation gets extended beyond tree level, where we recall it reads
\begin{equation}\label{eq:treelevelderivative}
    a_{n,\ell}^{(1)}=\frac12\frac\de{\de n}\left(a_{n,\ell}^{(0)}\ga_{n,\ell}^{(1)}\right)\,.
\end{equation}
At one loop, we can combine \eqref{eq:oneloopderivatives}, \eqref{eq:oneloopderivativet}, and its $u$-channel analog into a single relation,
\begin{equation}\label{eq:oneloopderivativeall}
    a_{n,\ell}^{(2)}=\frac12\frac\de{\de n}\left(a_{n,\ell}^{(0)}\gamma_{n,\ell}^{(2)}+a_{n,\ell}^{(1)}\ga_{n,\ell}^{(1)}\right)-\frac18\frac{\de^2}{\de n^2}\left(a_{n,\ell}^{(0)}\big(\ga_{n,\ell}^{(1)}\big)^2\right) \,.
\end{equation}
Identical relations hold for hatted quantities as defined in section \ref{sec:schemeindependent}.

It has been argued that a correction term in the derivative relations is necessary already at tree level for the exchange of single-trace operators \cite{Alday:2017vkk}, and a similar claim has been made for the quartic theory at one loop in $d=1$ \cite{Ferrero:2019luz}. Such corrections, however, depend on the choice of analytic continuation in $n$ for the OPE data, and are therefore not universal. In the cases where the full spectral function is known at a given perturbative order, one can choose a prescription that leads to derivative relations free of correction terms.
We recall that the spectral function $\rho(\uDi,\ell)$ is the spectral density of the four-point function $\calG(z,\bar{z})$,
\begin{equation}\label{eq:generalspectraldeco}
    \calG(z,\bar z)=1-\sum_\ell \int \frac{d\uDi}{2\pi i}\,\rho(\uDi,\ell)\,\calK_{\uDi,\ell}(z,\bar z)\,.
\end{equation}
The full non-perturbative spectral function has only simple poles corresponding to primaries $\calO$ in the $\phi\times \phi$ OPE,
\begin{equation}
    \rho(\uDi,\ell)=\sum_{\calO}\frac{a_{\calO}}{\uDi-\Di_\calO} + \text{ regular}\,,
\end{equation}
such that when closing the contour in \eqref{eq:generalspectraldeco} one recovers the block expansion \eqref{eq:genericblockexp}. Like $\calG(z,\bar z)$, also $\rho(\uDi,\ell)$ admits a perturbative expansion,
\begin{equation}
    \rho(\uDi,\ell)=\rho^{(0)}(\uDi,\ell)+\la\,\rho^{(1)}(\uDi,\ell)+\la^2\,\rho^{(2)}(\uDi,\ell)+\dots\,,
\end{equation}
with the leading term determined by MFT data, consisting of double-trace operators only,
\begin{equation}
    \rho^{(0)}(\uDi,\ell)=\sum_{n=0}^\infty\frac{a_{n,\ell}^{(0)}}{\uDi-\Di_{n,\ell}^{(0)}} + \text{ regular}\,.
\end{equation}
Similarly, the functions $\rho^{(k)}$ have poles at every double-trace location $\uDi = \Delta^{(0)}_{n,\ell}$, whose Laurent coefficients encode the CFT data at order $k$.
They can also have additional poles, corresponding to new operators contributing at that order, but let us focus for simplicity on the double-trace ones in what follows. For instance,
\begin{align}
    \rho^{(1)}(\uDi,\ell)&\underset{\uDi\sim \Di_{n,\ell}^{(0)}}{\sim}\frac{a_{n,\ell}^{(0)}\ga_{n,\ell}^{(1)}}{\big(\uDi-\Di_{n,\ell}^{(0)}\big)^2} +\frac{a_{n,\ell}^{(1)}}{\uDi-\Di_{n,\ell}^{(0)}} +\dots \,,\\
    \rho^{(2)}(\uDi,\ell)&\underset{\uDi\sim \Di_{n,\ell}^{(0)}}{\sim}\frac{a_{n,\ell}^{(0)}\big(\ga_{n,\ell}^{(1)}\big)^2}{\big(\uDi-\Di_{n,\ell}^{(0)}\big)^3} +\frac{a_{n,\ell}^{(0)}\ga_{n,\ell}^{(2)}+a_{n,\ell}^{(1)}\ga_{n,\ell}^{(1)}}{\big(\uDi-\Di_{n,\ell}^{(0)}\big)^2} +\frac{a_{n,\ell}^{(2)}}{\uDi-\Di_{n,\ell}^{(0)}} +\dots\,.
\end{align}
As an analytic function of $\uDi$, $\rho^{(k)}$ provides a way to fix the continuation of the OPE data to arbitrary $n$. In particular, correction terms to the derivative relations need not arise. 
Let us display the argument with a toy example. Consider the following $O(\lambda)$ contribution to the spectral function,
\begin{equation}
    \rho^{(1)}(\underline{\Delta}, \ell) = \frac{\pi^2}{4} \frac{F_\ell(\underline{\Delta})}{\sin^2\!\left(\frac{\pi}{2}(\underline{\Delta} - 2\Delta_\phi - \ell)\right)},
\end{equation}
with $F_\ell(\underline{\Delta})$ regular at the double-trace poles $\Delta_{n,\ell}^{(0)}$. Then we can extract the first correction to the CFT data as
\begin{equation}
    a_{n,\ell}^{(0)}\gamma_{n,\ell}^{(1)} = F_\ell\!\left(\Delta_{n,\ell}^{(0)}\right) \equiv A_{n,\ell}\,, \qquad a_{n,\ell}^{(1)} = \partial_{\underline{\Delta}} F_\ell(\underline{\Delta})\big|_{\underline{\Delta}=\Delta_{n,\ell}^{(0)}}\,.
\end{equation}
Now suppose $F_\ell(\underline{\Delta}) = 1 + \sin\!\left(\tfrac{\pi}{2}(\underline{\Delta} - 2\Delta_\phi - \ell)\right)$, so that $A_{n,\ell} \equiv 1$ and $a_{n,\ell}^{(1)} = \tfrac{\pi}{2}(-1)^n$. Since $A_{n,\ell}$ is constant in $n$, one might conclude that a correction term in the derivative relation is necessary,
\begin{equation}
    a_{n,\ell}^{(1)} = \tfrac{1}{2}\partial_n A_{n,\ell} + B_{n,\ell}\,, \qquad B_{n,\ell} = \tfrac{\pi}{2}(-1)^n\,.
\end{equation}
A more convenient analytic extension of $A_{n,\ell}$ is however $F_\ell(2\Delta_\phi + 2n + \ell) = 1 + \sin(\pi n)$, which reproduces the correct $a_{n,\ell}^{(1)}$ with $B_{n,\ell} = 0$. 

Having illustrated the ambiguous nature of the correction terms, let us see how relations like \eqref{eq:treelevelderivative} and \eqref{eq:oneloopderivativeall} arise in the spectral representation in full generality. To do so, let us discuss the pole structure of $\rho^{(k)}(\uDi,\ell)$. In general, at double-trace locations it will have poles of order at most $p_{k,\ell}+1$, with $1\leq p_{k,\ell}\leq k$, or it will be regular, in which case we set $p_{k,\ell}=-1$. It is then convenient to split $\rho^{(k)}$ as follows,
\begin{equation}
    \rho^{(k)}(\uDi,\ell)=\calM^{(k)}(\uDi,\ell)\calR^{(k)}(\uDi,\ell)\,,
\end{equation}
with $\calR^{(k)}(\uDi,\ell)$ being an analytic function of $\uDi$ which is regular near the poles, and $\calM^{(k)}$ being a meromorphic function that captures the leading divergence at each pole,
\begin{equation}
    \calM^{(k)}(\uDi,\ell)\underset{\uDi\sim \Di_{n,\ell}^{(0)}}\sim \frac1{\big(\uDi-\Di_{n,\ell}^{(0)}\big)^{p_{k,\ell}+1}}+\text{ regular}\,,\qquad \forall\, n\in \ZZ_{\geq0}\,,
\end{equation}
in particular it has no subleading poles. A function that satisfies such requirements is
\begin{equation}
    \calM^{(k)}(\uDi,\ell)=\sum_{n=0}^\infty \frac1{\big(\uDi-\Di_{n,\ell}^{(0)}\big)^{p_{k,\ell}+1}}=\frac1{2^{p_{k,\ell}+1}p_{k,\ell}!}\psi^{(p_{k,\ell})}\left(\frac{2\Delta_\phi+\ell-\uDi}{2}\right)\,.
\end{equation}
We intentionally avoid introducing an explicit $n$ dependence, so that the analytic $n$ dependence only enters through $\uDi\to \Di_{n,\ell}^{(0)}$, making the identification $\de_{\uDi}\to \frac12\de_n$ unambiguous. When closing the contour in \eqref{eq:generalspectraldeco}, we then have
\begin{equation}
\begin{aligned}
     \calG^{(k)}(z,\bar z)&\supset \sum_{n,\ell}\Res_{\uDi=\Di_{n,\ell}^{(0)}}\left[\rho^{(k)}(\uDi,\ell)\calK_{\uDi,\ell}(z,\bar z)\right]\\
     &=\sum_{n,\ell}\left.\frac1{p_{k,\ell}!}\frac{\de^{p_{k,\ell}}}{\de\uDi^{p_{k,\ell}}}\left[\calR^{(k)}(\uDi,\ell)\calK_{\uDi,\ell}(z,\bar z)\right]\right|_{\uDi=\Di_{n,\ell}^{(0)}}\\
     &=\sum_{n,\ell}\sum_{r=0}^{p_{k,\ell}}\left.\left[A_r^{(k)}(\uDi,\ell)\partial_{\uDi}^r\calK_{\uDi,\ell}(z,\bar z)\right]\right|_{\uDi=\Di_{n,\ell}^{(0)}}\,,
\end{aligned}
\end{equation}
where the functions $A_r^{(k)}$ are obtained by expanding the derivative,
\begin{equation}
    A_r^{(k)}(\uDi,\ell)=\frac1{r!(p_{k,\ell}-r)!}\partial_{\uDi}^{p_{k,\ell}-r}\calR^{(k)}\,.
\end{equation}
The functions $A_r^{(k)}$ manifestly satisfy derivative identities among themselves. For instance, at $k=1$, we have $p_{1,0}=1$ and $p_{1,\ell>0}=-1$, so the only nontrivial relation is
\begin{equation}
    A_0^{(1)}(\uDi,0)=\partial_{\uDi} A_1^{(1)}(\uDi,0)\,,
\end{equation}
which, once we set $\uDi=\Di_{n,0}^{(0)}$, leads to \eqref{eq:treelevelderivative}. At $k=2$, instead, we have $p_{2,0}=2$ and $p_{2,\ell>0}=1$, so we get three basic relations,
\begin{equation}\label{eq:alloneloopderivativerelations}
    \begin{cases}
        A_0^{(2)}(\uDi,\ell>0)=\de_{\uDi}A_1^{(2)}(\uDi,\ell>0)\,,\\
        A_0^{(2)}(\uDi,0)= \frac12 \de_{\uDi}A_1^{(2)}(\uDi,0)\,,\\
        A_1^{(2)}(\uDi,0)=2\de_{\uDi} A_2^{(2)}(\uDi,0)\,.
    \end{cases}
\end{equation}
The derivative relation \eqref{eq:oneloopderivativeall} is then a special instance of the identities for $A_r^{(2)}$ upon setting $\uDi=\Di_{n,\ell}^{(0)}$, namely
\begin{equation}
    A_0^{(2)}(\uDi,\ell)=\de_{\uDi}A_1^{(2)}(\uDi,\ell)-\de_{\uDi}^2A_2^{(2)}(\uDi,\ell)\,,
\end{equation}
and this is the only combination that is valid for all spins simultaneously, with $A_2^{(2)}(\uDi,\ell>0)\equiv0$. 

We should also note that the third relation in \eqref{eq:alloneloopderivativerelations}, which connects $A^{(2)}_1$ to $A^{(2)}_2$, might naively suggest that the one-loop CFT data can be fixed from tree-level data alone. This is not the case: extracting this relation required knowledge of the full one-loop spectral function. In particular, evaluating $\partial_\Delta A^{(2)}_2$ at the double-trace poles requires knowing the full analytic function $A_2^{(2)}$ near the poles; this analytic function is fixed by the one-loop spectral function $\rho^{(2)}$, not by the tree-level data as determined by $\rho^{(1)}$. Both the analytic continuation of the CFT data and the form of the corresponding derivative relations also depend on the choice of split $\rho^{(k)}=\mathcal{M}^{(k)}\mathcal{R}^{(k)}$. Our prescription puts only the leading pole at each double-trace location into $\mathcal{M}^{(k)}$; if subleading pole terms are absorbed into $\mathcal{M}^{(k)}$ instead, the relations \eqref{eq:alloneloopderivativerelations} acquire additional terms proportional to the corresponding Laurent coefficients.

We have seen that, with this prescription, derivative relations such as \eqref{eq:treelevelderivative} and \eqref{eq:oneloopderivativeall} are automatically built into the spectral representation approach, and reduce to elementary statements about residues. The correction terms found in \cite{Alday:2017vkk,Ferrero:2019luz} are of course correct in their respective contexts, where anomalous dimensions are determined at integer $n$ and a choice of analytic continuation must be made independently; the correction terms measure the discrepancy between that choice and the continuation defined here by the full spectral function together with our prescription for $\mathcal{M}^{(k)}$. A direct comparison between the two approaches would provide an independent validation of the analytic continuation used here and a useful cross-check of our results; we leave this for future work.

\subsection{Monotonicity properties in spin}
\label{subsec:monotonicity}

The $t$-channel contribution to the anomalous dimension can be written as $\gamma^{(2,t)}_{n,\ell} = (-1)^{\ell+1}\,E_{n,\ell}$, where $E_{n,\ell} > 0$ is a smooth positive function of $\ell > 0$.
Since the $s$-channel contributes only at $\ell = 0$,  the full anomalous dimension at $\ell>0$, $\ell$ even, is 
\begin{equation}
\gamma^{(2)}_{n,\ell} = -2\,E_{n,\ell} < 0\,.
\end{equation}
The function $E_{n,\ell}$ is actually a
completely monotonic function of $\ell$,
\begin{equation}
(-1)^r\,\frac{\de^r}{\de\ell^r}\,E_{n,\ell} > 0\,.
\end{equation}
We prove this as follows. Expanding the
${}_{7}W_6$ in $E_{n,\ell}$ as $\sum_{k}H_{k,q}(m)$, where $m=n+\ell$, each term factorizes as
\begin{equation}
\Lambda_{m,q}\,H_{k,q}(m) = B_{k,q} G_k(m) R_{k,q}(m)\,,
\end{equation}
with $B_{k,q}= \frac{\Gamma(2\Di_\phi-1)}{2\Di_\phi-1} \frac{(2\Di_\phi-1)_k(\Di_\phi+q)^2_k}{(\Di_\phi+1/2)_k} > 0$, and
\begin{equation}
G_k(m)
= \frac{\Gamma(m+1)}{\Gamma(m+2\Delta_\phi+k)}
= \frac{1}{\Gamma(2\Delta_\phi+k-1)}
  \int_0^1 dt\, t^m (1-t)^{2\Delta_\phi+k-2}\,,
\end{equation}
which leads to
\begin{equation}
(-1)^r\frac{\de^r}{\de m^r}G_k(m)
= \frac{1}{\Gamma(2\Delta_\phi+k-1)}
  \int_0^1 dt\, |\log t|^r\, t^m (1-t)^{2\Delta_\phi+k-2} > 0\,.
\end{equation} 
The remaining factor
$$R_{k,q}(m)=\frac{m+3\Di_\phi-1+q+2k}{(m+\Di_\phi+q)(m+2\Di_\phi-1)}\frac{(m+2\Di_\phi+q-1/2)_k}{(m+\Di_\phi+q+1)_k(m+2\Di_\phi)_k}\,,$$
decomposes into a product of elementary fractions of the form $\frac{m + \chi}{(m+\alpha)(m+\beta)}$,
each of which is completely monotonic whenever $\chi \geq \min(\alpha,\beta)$, a condition verified by inspection. Since products of completely monotonic functions are completely monotonic, each term $\Lambda_{m,q} H_{k,q}(m)$ is
completely monotonic in $m$. Very-well-poisedness of the $_7W_6$ guarantees that the series and all its $m$ derivatives converge locally uniformly,
justifying termwise differentiation and hence the complete monotonicity of $E_{n,\ell}$. We conclude that the sign pattern of all $\ell$ derivatives of
$\gamma^{(2)}_{n,\ell}$ at even $\ell>0$ is
\begin{equation}
\label{eq:sign_pattern}
\gamma^{(2)}_{n,\ell} < 0, \qquad
\partial_\ell\gamma^{(2)}_{n,\ell} > 0, \qquad
\partial_\ell^2\gamma^{(2)}_{n,\ell} < 0, \qquad
\partial_\ell^3\gamma^{(2)}_{n,\ell} > 0, \qquad \ldots\,.
\end{equation}

Negativity and monotonicity at $n=0$ were argued in \cite{Aharony:2016dwx} using Nachtmann's theorem and the lightcone bootstrap. Since
$\gamma^{(1)}_{0,\ell>0}=0$, the one-loop anomalous dimensions
are the leading correction to mean field theory for all spinning double-trace operators. Negativity follows from large-spin asymptotics: the leading term
$\gamma^{(2)}_{0,\ell}\sim -c/{\sf J}^{2\Delta_\phi}$, with $c>0$, is manifestly negative, and monotonicity follows from reflection positivity of the $\phi\times\phi$ OPE and the even-spin
constraint \cite{Fitzpatrick:2012yx,Komargodski:2012ek}. 

Convexity of the twist spectrum was established in \cite{Komargodski:2012ek} via deep inelastic scattering in a gapped deformation of the CFT, and the lightcone bootstrap was used there to show that the leading large-spin correction to double-trace anomalous dimensions is negative, with its sign governed by the positivity of OPE coefficients of minimal-twist $t$-channel operators. In the bulk, $-\gamma_{n,\ell}$
is proportional to the two-particle scattering phase at impact parameter $b\sim\log\ell$, which must be positive by causality \cite{Camanho:2014apa,Cornalba:2006xm,Cornalba:2007zb}. Negativity, monotonicity, and
convexity were also verified for specific tree-level single-trace exchanges
in \cite{Alday:2017vkk}. Although in our case the boundary theory does not have a gravitational
dual, the properties of negativity, monotonicity, and convexity established in \cite{Komargodski:2012ek} extend here to an alternating-sign pattern for all derivatives with respect to spin.

The complete monotonicity of $E_{n,\ell}$ is equivalent to it being the Laplace transform of a non-negative function $\mathcal{E}_n(s)$,
\begin{equation}
    E_{n,\ell} = \int_0^\infty ds\, e^{-s\ell}\,\mathcal{E}_n(s)\,, \qquad \mathcal{E}_n(s) \geq 0\,.
\end{equation}
The large-spin asymptotics, controlled by the minimal twist $2\Delta_\phi$ of the $t$-channel exchanged operators, then fixes the behavior of $\mathcal{E}_n(s)$ near $s=0$ by a Tauberian theorem,
\begin{equation}
    \mathcal{E}_n(s) \underset{s\to 0^+}\sim \frac{c_n}{\Gamma(2\Delta_\phi)}\,s^{2\Delta_\phi-1}\,.
\end{equation}
This argument is structurally identical to the one used in \cite{Qiao:2017xif} to put the lightcone bootstrap on rigorous footing, with the positivity of $\mathcal{E}_n$ playing the role of positivity of OPE coefficients, and the large-spin expansion in place of the large-$\Delta$ expansion. Given the parallels, it is tempting to speculate whether the complete monotonicity of anomalous dimensions is a general consequence of unitarity. This would be consistent with the broader observation that complete monotonicity appears to be a pervasive property of QFT observables \cite{Henn:2024qwe}. It would be interesting to formalize this further.

\section*{Acknowledgments}
We thank Ivo Sachs and Weichen Xiao for discussions, as well as Agnese Bissi and Lorenzo Di Pietro for comments on the draft. The work of the authors is supported by the Israel Science Foundation (ISF) Grant No. 1487/21, by the MOST NSF/BSF physics grant number 2022726, and by the BSF grant number 2024187. The work of S.S. is also partially supported
by a PhD fellowship from the Israel Scholarship Education Foundation.

\appendix

\section{Sketch of the proof of eq. \eqref{eq:finitegamma2t} } \label{sec:proof}
The starting point is \eqref{eq:infinitegamma2t}, which, after making everything explicit, becomes
\begin{equation}
\begin{aligned}
\gamma_{n,\ell}^{(2,t)}(\Delta_\phi)
&=
\frac{(-1)^{\ell+1}}{64\pi^2}\,
\frac{\Gamma(m+1)}{\Gamma(n+1)}\,
\frac{
(\Delta_\phi-n)_n\,
(n+2\Delta_\phi-1)_n
}{
(\Delta_\phi)_n\,
(m+\Delta_\phi)\,
(m+2\Delta_\phi-1)^2\,
(2\Delta_\phi-1)_m
}
\\
&\quad\times
\sum_{p=0}^{\infty}
\frac{
(n+\Delta_\phi)_p\,
(2\Delta_\phi-1)_p\,
(\Delta_\phi)_p^2\,
(2\Delta_\phi+2m+1)_p
}{
(\Delta_\phi-n)_p\,
(m+2\Delta_\phi)_p^2\,
(m+\Delta_\phi+1)_p^2
}
\\
&\quad\times
{}_4F_3\!\left(
\begin{matrix}
-n,\,-n,\,1-n-\Delta_\phi,\,1-n-\Delta_\phi\\
2-2n-2\Delta_\phi,\,1-n-p-\Delta_\phi,\,p-n+\Delta_\phi
\end{matrix}
;1
\right)
\\
&\quad\times
{}_7W_6\!\left(
2\Delta_\phi+2m+p;\begin{matrix}
    m+1,\,m+1,\,p+1,\\
m+\Delta_\phi,\,m+\Delta_\phi
\end{matrix}
\right)\,.
\end{aligned}
\end{equation}

The first step is to notice that the ${}_4F_3(1)$ in \eqref{eq:smallomega} admits the finite expansion
\begin{equation}
{}_4 F_3\!\left(
\begin{matrix}
-n,-n,1-\Delta_\phi-n,1-\Delta_\phi-n\\
2-2\Delta_\phi-2n,1-\Delta_\phi-n-p,\Delta_\phi-n+p
\end{matrix};1\right)
=
\mathcal N_{n,p}
\sum_{r=0}^n
\alpha_{n,r}
\frac{(\Delta_\phi+r)_p}{(\Delta_\phi-r)_p}\,,
\end{equation}
with
\begin{equation}
\alpha_{n,r}
=
\frac{(-n)_r(n+2\Delta_\phi-1)_r(1-\Delta_\phi)_r}
{(1)_r(\Delta_\phi)_r r!}\,.
\end{equation}
After absorbing the prefactor $\mathcal N_{n,p}$ and the remaining prefactors into the summand, the anomalous dimension becomes
\begin{equation}
\gamma_{n,\ell}^{(2,t)}(\Delta_\phi)
=
\frac{(-1)^{\ell+1}}{128\pi^2}
\sum_{r=0}^n \alpha_{n,r}\,I_r \,.
\end{equation}
The quantity $I_r$ is most conveniently analyzed by decomposing the
very-well-poised ${}_7F_6$ appearing in $\widehat\Omega_\phi(\uDi_p,\Di_\phi+m)$, $m=n+\ell$, back into two ${}_4F_3$'s,
\begin{equation}
\begin{aligned}
    {}_7W_6&\left(2\Di_\phi+2m+p;\begin{array}{c}
         1+m,1+m,1+p  \\
          \Di_\phi+m,\Di_\phi+m
    \end{array}\right)\\
&=
A^{(1)}(p)\,
{}_4F_3\!\left(
\begin{matrix}
1+p,\ \Delta_\phi+p,\ \Delta_\phi+p,\ 2\Delta_\phi+p-1
\\ 
2\Delta_\phi+2p,\ m+2\Delta_\phi+p,\ 1-m+p
\end{matrix}
;1
\right)\\& 
+
A^{(2)}(p)\,
{}_4F_3\!\left(
\begin{matrix}
m+1,\ m+\Delta_\phi,\ m+\Delta_\phi,\ m+2\Delta_\phi-1
\\ 
m+2\Delta_\phi+p,\ 2m+2\Delta_\phi,\ 1+m-p
\end{matrix}
;1
\right)\,,
\end{aligned}
\end{equation}
where $A^{(i)}(p)$ are the appropriate prefactors. Expanding the $_4F_3$'s and rearranging the summation in each term we arrive at
\begin{equation}
I_r=I_r^{(1)}+I_r^{(2)}\,,
\end{equation}
\begin{equation}
I_r^{(1)}
=
P_m\sum_{p=0}^{\infty}
\frac{A_r(p)}{\Gamma(1+p-m)}\,,
\qquad
I_r^{(2)}
=
-P_m\sum_{p=0}^{\infty}
\frac{A_r(p+m)}{\Gamma(1+p)}\,,
\end{equation}
where
\begin{equation}
P_m=
\frac{\Gamma(2\Delta_\phi-1)^2}{\Gamma(2\Delta_\phi)\Gamma(2\Delta_\phi+m)}
\Gamma(m)\Gamma(1-m)\,,
\end{equation}
and
\begin{equation}
A_r(x)=
\frac{
\left(\frac12\right)_x(\Delta_\phi)_x(2\Delta_\phi-1)_x}
{\left(\Delta_\phi+\frac12\right)_x(2\Delta_\phi+m)_x}
\frac{
\left(\frac12\right)_r(1-\Delta_\phi-x)_r}
{\left(\frac12-x\right)_r(1-\Delta_\phi)_r}\,.
\end{equation}
The two terms are separately singular when $m\in\mathbb Z_{\geq 0}$, but their sum is regular. 
The tempting direct truncation of the difference of the two sums is not
legitimate before the integer $m$ limit is regularized. We
therefore change basis before taking this limit. Introduce the triangular kernel
\begin{equation}
K_{q,r}:=
\frac{q!\left(\frac12\right)_q}
{(\Delta_\phi)_q\left(\Delta_\phi-\frac12\right)_q}
\frac{
(-q)_r(1-\Delta_\phi)_r\left(\Delta_\phi+q-\frac12\right)_r}
{(1)_r\left(\frac12\right)_r r!}\,,
\qquad 0\leq r\leq q \,.
\end{equation}
It is chosen so that
\begin{equation}
B_q(x)\equiv\sum_{r=0}^q K_{q,r}A_r(x)
\end{equation}
can be summed in closed form. Explicitly,
\begin{equation}
B_q(x)
=
\frac{
\left(\frac12\right)_x(\Delta_\phi)_x(2\Delta_\phi-1)_x}
{\left(\Delta_\phi+\frac12\right)_x(2\Delta_\phi+m)_x}
\frac{
\left(\frac12\right)_q(\Delta_\phi+x)_q}
{(\Delta_\phi)_q\left(\frac12-x\right)_q}\,.
\end{equation}
Moreover, the regulated combination
\begin{equation}
J_q\equiv\sum_{r=0}^q K_{q,r}I_r
\end{equation}
can be resummed to a very-well-poised $_7F_6(1)$. One finds
\begin{equation}
J_q
=
\Lambda_{m,q}
\,
{}_7W_6\!\left(
m+3\Delta_\phi+q-1;\begin{array}{c}
    2\Delta_\phi-1,\Delta_\phi+q,\Delta_\phi+q,   \\
      m+2\Delta_\phi+q-\frac12,1
\end{array}
\right)\,.
\end{equation}
This expression is manifestly regular at non-negative integer $m$.

It remains to express the original coefficients $\alpha_r$ in the
triangular basis $K_{q,r}$. Namely, we solve
\begin{equation}
\alpha_{n,r}
=
\sum_{q=r}^n  c_{n,q}K_{q,r}\,.
\end{equation}
The triangular inversion gives
\begin{equation}
c_{n,q}
=\frac{\left(\Delta_\phi-\frac12\right)_n}
{\left(\Delta_\phi+\frac12\right)_n}
\,
\frac{
(-n)_q
\left(n+2\Delta_\phi-1\right)_q
\left(\Delta_\phi-\frac12\right)_q^2
\left(\frac{\Delta_\phi}{2}+\frac34\right)_q
}{
q!\,(1)_q
\left(\frac32-\Delta_\phi-n\right)_q
\left(n+\Delta_\phi+\frac12\right)_q
\left(\frac{\Delta_\phi}{2}-\frac14\right)_q
}\,.
\end{equation}
Consequently,
\begin{equation}
\sum_{r=0}^n \alpha_{n,r} I_r
=
\sum_{q=0}^n  c_{n,q} J_q \,.
\end{equation}
Combining this with the overall prefactor gives the desired finite representation \eqref{eq:finitegamma2t}.

\bibliographystyle{JHEP}
\bibliography{biblio.bib}
\end{document}